\documentclass[bibyear]{aa}  
\usepackage{graphicx}
\usepackage[]{natbib}       
\usepackage{txfonts}
\usepackage{graphicx}	
\usepackage{amsmath}	
\usepackage{amssymb}
\usepackage{upgreek}
\usepackage{ulem}
\usepackage{xcolor}
\usepackage{algpseudocode}
\usepackage{hyperref}
\newcommand{\msol}{\mathrm{M_{\odot}}}
\newcommand{\kgmcube}{\mathrm{kg\,m^{-3}}}

\newcommand{\jmsqare}{\mathrm{J\,m^{-2}}}

\newcommand{\msec}{\mathrm{m\,s^{-1}}}

\begin{document}

   \title{Compaction during fragmentation and bouncing produces realistic dust grain porosities in protoplanetary discs}
   \titlerunning{Realistic porosities via dust compaction}
   

   \author{Stéphane Michoulier
          \inst{1}
          \and
          Jean-François Gonzalez\inst{1}
          \and
          Daniel J. Price\inst{2}
        }

   \institute{Universite Claude Bernard Lyon 1, CRAL UMR5574, ENS de Lyon, CNRS, Villeurbanne, F-69622, France\\
              \email{jean-francois.gonzalez@ens-lyon.fr}
        \and
             School of Physics and Astronomy, Monash University, Vic. 3800, Australia
             }

   \date{Received 26 February 2024; accepted 21 June 2024}

 
  \abstract
   {In protoplanetary discs, micron-sized dust grows to form millimetre- to centimetre-sized pebbles but encounters several barriers during its evolution. Collisional fragmentation and radial drift impede further dust growth to planetesimal size. Fluffy grains have been hypothesised to solve these problems. While porosity leads to faster grain growth, the implied porosity values obtained from previous simulations were larger than suggested by observations.}
   { In this paper, we study the influence of porosity on dust evolution taking into account growth, bouncing, fragmentation, compaction, rotational disruption and snow lines, in order
   to understand their impact on dust evolution. }
   {We develop a module for porosity evolution for the 3D Smoothed Particle Hydrodynamics (SPH) code \textsc{Phantom} that accounts for dust growth and fragmentation. This mono-disperse model is integrated into both a 1D code and the 3D code to capture the overall evolution of dust and gas.}
   {We show that porosity helps dust growth and leads to the formation of larger solids than when considering compact grains, as predicted by previous work. Our simulations taking into account compaction during fragmentation show that large millimetre grains are still formed, but are 10 to 100 times more compact. Thus, mm sizes with typical filling factors of $\sim0.1$ match the values measured on comets or via polarimetric observations of protoplanetary discs. 
   }
  {}
   \keywords{methods: numerical --- planets and satellites: formation --- protoplanetary discs}

   \maketitle
%

\section{Introduction}
The early stages of planet formation are not well understood. 
Observations of protoplanetary and debris discs, along with the discovery of numerous exoplanets, have shed light on the efficiency of mechanisms that allow tiny dust particles to aggregate into objects several thousand kilometres in size \citep{dominik_physics_1997,Dullemond_2005}.
The process of dust particle growth involves several stages. 
Initially, monomers grow thanks to Brownian motions, sticking together via Van der Waals forces \citep{Cuzzi_1993,stepinski_global_1997}. 
However, after growing by just one order of magnitude, this mechanism becomes ineffective, giving way to turbulence within the gas, which promotes collisions between aggregates \citep{weidenschilling_formation_1993}. 
Thanks to this turbulence, grains can grow to sizes ranging from millimetres to metres, depending on their porosity \citep{okuzumi_rapid_2012,garcia_evolution_2020}. At this point, they face various barriers to further growth, including radial drift and fragmentation.

Radial drift is caused by the friction between gas and dust \citep{whipple_certain_1972,weidenschilling_aerodynamics_1977}, resulting from the difference in orbital velocities due to gas pressure support. This friction causes dust particles to lose angular momentum and drift inward towards the star \citep{nakagawa_settling_1986}. This phenomenon is especially rapid for intermediate-sized grains, posing a challenge to their growth \citep{weidenschilling_aerodynamics_1977}.
In addition to radial drift, other barriers to dust growth are associated with the physics of grain interactions \citep{weidenschilling_formation_1993,stepinski_global_1997}. 
These interactions depend on the relative velocities of grains during collisions, which can lead to coagulation, bouncing (elastic or plastic deformation), or fragmentation \citep{blum_experiments_2000}. 
The bouncing or fragmentation thresholds refer to the relative velocity at which grains begin to bounce or fragment, respectively. In any case, these thresholds, ranging from a few to a few tens of metres per second, are quickly reached when sizes are on the order of millimetres to metres. These barriers are called the bouncing barrier \citep{Zsom_2010, Windmark_2012} and the fragmentation barrier \citep{weidenschilling_formation_1993, dominik_physics_1997, blum_growth_2008}.
\citet{blum_experiments_2000,blum_growth_2008} and \citet{guttler_outcome_2010} showed in their experiments that adhesion and fragmentation are linked to the composition of the grains and the volatile materials covering them.
Recent studies \citep{yamamoto_examination_2014, kimura_tensile_2020, san_sebastian_tensile_2020} show that silicates are more resistant than suspected.

More recently, additional barriers to dust growth have been identified, including collisional erosion \citep{Schrapler_2011} and aeolian erosion \citep{Paraskov_2006_erosion, Rozner_2020_Erosiona, Grishin_2020_Erosionb, Michoulier_erosion}. 
Collisional erosion occurs when larger aggregates eject grains from their surface through successive collisions with smaller grains. 
Aeolian erosion is caused by gas friction, leading to the detachment of loosely bound grains from larger aggregates in the innermost region of the disc.
Another mechanism that can destroy grains is the rotation disruption \citep{tatsuuma_rotational_2021}, where the gas flow torques the grains, causing them to rotate. 
When the centrifugal force exceeds the maximum tensile strength of the grains, they shatter into fragments. \citet{tatsuuma_rotational_2021} and  \citet{Michoulier_Gonzalez_Disruption} studied this process for porous aggregates in protoplanetary discs.

All of these barriers theoretically hinder dust growth. In order to understand how the barriers are bypassed in nature to form planetesimals from micron-sized grains, various solutions have been proposed. Some of them involve creating dust traps with local pressure maxima, such as vortices \citep{barge_did_1995,meheut_formation_2012}, 
the baroclinic instability \citep{Klahr_2003,loren-aguilar_toroidal_2015}, planet gaps \citep{Paardekooper_2004,Fouchet_2007,Fouchet_2010,Gonzalez_2012,Zhu_2014} or snow lines where different chemical species sublimate \citep{kretke_grain_2007,brauer_planetesimal_2008,drazkowska_modeling_2014,drazkowska_planetesimal_2017,Hyodo_2019,vericel_self-induced_2020}. 
Self-induced dust traps, driven by dust feedback, have also been proposed by \citet{gonzalez_self-induced_2017}. Other mechanisms enable direct formation of planetesimals, like the streaming instability \citep{youdin_streaming_2005,Johansen_2007, youdin_protoplanetary_2007, Jacquet_2011, Carrera_2015, yang_concentrating_2017, schafer_initial_2017,auffinger_linear_2018, li_demographics_2019,Zhu_2021, Schaffer_2021}. The streaming instability arises from the weak coupling of large dust grains and gas in dust enriched regions, and concentrates dust in filamentary structures, which can ultimately collapse gravitationally.

One solution that has gained recent attention is grain porosity \citep{ormel_dust_2007, suyama_numerical_2008, okuzumi_numerical_2009, suyama_geometric_2012, okuzumi_rapid_2012, kataoka_fluffy_2013, garcia_evolution_2018, garcia_evolution_2020}. 
Porosity, often overlooked for simplicity in dust modelling \citep{weidenschilling_aerodynamics_1977, nakagawa_settling_1986, barriere-fouchet_dust_2005, laibe_growth_2008, drazkowska_modeling_2014, gonzalez_accumulation_2015, vericel_dust_2021}, changes gas-dust coupling and grain evolution. 
Porous grains, due to their larger collision cross-sections for a given mass, grow faster and decouple from the gas at larger sizes, enhancing their survival in the disc. This has been observed in studies using local or 1D disc models to track the evolution of mass and filling factor for single grains \citep{ormel_dust_2007, suyama_numerical_2008, okuzumi_numerical_2009, suyama_geometric_2012, okuzumi_rapid_2012, kataoka_fluffy_2013} and more recently in a 3D disc model evolving an entire population of porous grains \citep{garcia_evolution_2018, garcia_evolution_2020}.
They are also less susceptible to fragmentation, contributing to planetesimal formation through coagulation \citep{garcia_evolution_2018}.
Laboratory experiments studying growth of highly porous aggregates are restricted to low velocity \citep{Blum_2004}, making numerical experiments necessary \citep{blum_experiments_2000, dominik_physics_1997, Wada_2007, suyama_numerical_2008, wada_collisional_2009, suyama_geometric_2012, Seizinger_2012}. 

\citet{guttler_synthesis_2019}, by looking at dust in comets, classified aggregates into three classes: fractal, porous, and compact. They also showed that porous grains are common, supporting the need to account for porosity in models. 
Observations of discs show that porosity must be considered to explain spectral energy distributions \citep[SEDs, ][]{Kataoka_polar_2016,Zhang_2023}, polarisation \citep{Kataoka_polar_2015, Kataoka_polar_2016, Kataoka_2019_polar, Tazaki_kataoka_2019} and the amount of dust settling \citep{Pinte_2019,Verrios_2022}. 
The monomer size is key to explaining polarized light \citep{tazaki_how_2022}, and smaller grains on disc surfaces are identified as fractal or porous aggregates \citep{Tazaki_monomers_2023}. 

In this paper, we focus on the impact of porosity on dust evolution by modelling mechanisms impairing the growth of grains, in the frame of the mono-disperse approximation. Sections~\ref{Sc:Methods} and \ref{Sc:Simulations} present our model and numerical simulations, respectively.
In Sect.~\ref{Sc:Results}, we show the impact of porosity on dust evolution and study the role of compaction during fragmentation, using 1D and 3D simulations.
We discuss our results and implications, as well as limitations of our models, in Sect.~\ref{Sc:Discussion} and conclude in Sect.~\ref{Sc:Conclusions}.

\section{Dust evolution model}\label{Sc:Methods}

\subsection{Grain growth}\label{Ssc:Growth Model}
Dust coagulation results from collisions between grains with a certain relative velocity. Different velocities or mass ratios can result in sticking, mass transfer or penetration \citep{guttler_outcome_2010}. To treat dust growth, \citet{stepinski_global_1997} assume that locally, the mass distribution of grains is peaked around a single value, or `mono-disperse', where collisions occur between grains of identical mass.
In the following, we adopt their formalism, following the implementations by \citet{laibe_growth_2008} or, for the \textsc{Phantom} code, \citet{vericel_dust_2021}. The reader is referred to these two papers for discussion of its implications.
In each collision, the mass $m$ of a dust grain of size $s$ doubles over a characteristic time $\tau_\mathrm{coll}$
\begin{equation}\label{Eq:Growth_dmdt_Stepinski}
    \left(\frac{\mathrm{d} m}{\mathrm{d} t}\right)_{\mathrm{grow}} \approx \frac{m}{\tau_{\mathrm{coll}}}.
\end{equation}
The characteristic time can be expressed in terms of the dust number density 
$n_\mathrm{d}$, collision cross-sectional area ($\sigma_\mathrm{d} = 4 \pi s^2$), 
and relative velocity $v_\mathrm{rel}$ during collision as $\tau_\mathrm{coll} = \left(n_\mathrm{d} \sigma v_\mathrm{rel}\right)^{-1}$.
The mass change rate can then be computed as

\begin{equation}\label{Eq:Growth_dmdt_Stepinski_bis}
    \left(\frac{\mathrm{d} m}{\mathrm{d} t}\right)_{\mathrm{grow}}= 4 \pi \rho_{\mathrm{d}} s^2 v_{\mathrm{rel}},
\end{equation}
where $\rho_\mathrm{d}$ is the dust volume density, while the relative velocity $v_\mathrm{rel}$ can be expressed as
\begin{equation}\label{Eq:Vrel}
    v_{\mathrm{rel}} = \sqrt{2} 
    v_{\mathrm{t}}\frac{\sqrt{\mathrm{Sc}-1}}{\mathrm{Sc}}.
\end{equation}
The turbulent velocity $v_{\mathrm{t}}$ is given by
\begin{equation}\label{Eq:Vt_definition}
    v_{\mathrm{t}} = \sqrt{2^{1/2}\mathrm{Ro}\, \alpha}\, c_{\mathrm{g}},
\end{equation}
$\mathrm{Ro}$ is the Rossby number, which is a constant equal to 3 \citep{stepinski_global_1997}, 
$\alpha$ the turbulent viscosity parameter \citep{shakura_black_1973} and $c_\mathrm{g}$ the sound speed.
The Schmidt number $\mathrm{Sc}$ of a dust grain measures the grain's coupling to the vortex
\begin{equation}\label{Eq:Schmidt_definition}
    \mathrm{Sc} = \left(1+\mathrm{St}\right) \sqrt{1+\frac{\Delta v ^2}{v_{\mathrm{t}}^2}},
\end{equation}
where $\Delta v$ is the difference in velocity between grains due to friction with the gas. 
St measures the coupling between gas and dust, and is expressed as
\begin{equation}
   \mathrm{St} = t_\mathrm{s} \Omega_\mathrm{K}.
\end{equation}
where $t_\mathrm{s}$ is the stopping time, that is the time needed for a grain to reach the gas velocity, and $\Omega_\mathrm{K}$ the Keplerian frequency.
More details about the implementation and the growth model are provided in \citet{vericel_dust_2021}.

\subsection{Porosity evolution during growth}\label{Ssc:Porosity Model}

In this section, we describe the modelling of porosity evolution during the growth of an aggregate.
\citet{suyama_numerical_2008}, \citet{suyama_geometric_2012}, and \citet{okuzumi_rapid_2012} have developed a porosity evolution model. 
\citet{suyama_numerical_2008} and \citet{okuzumi_rapid_2012} use the Smoluchowski equation to evolve an aggregate, 
which makes the porosity model discrete --- the filling factor computed at iteration $n$ directly depends on 
the quantities from iteration $n-1$. While this allows for a detailed modelling of porosity evolution in an $N$-body simulation 
of grain collisions, the model cannot be used directly in global 3D simulations involving dust growth that rely on other 
methods. The continuous model developed by \citet{garcia_evolution_2018} enabled its implementation in the \textsc{LyonSPH} code 
\citep{barriere-fouchet_dust_2005, laibe_growth_2008, garcia_evolution_2020}.

An aggregate is a collection of several monomers, which we consider as compact spheres with mass $m_0$, size $a_0$, and intrinsic density $\rho_\mathrm{s}$.
Experiments \citep{blum_experiments_2000,shimaki_experimental_2012}, numerical simulations \citep{Seizinger_2012,Ringl_Bringa_2012,wada_growth_2013,Planes_2023}, 
and observations \citep{guttler_synthesis_2019} show that dust grains are not perfect spheres but rather have ovoid or irregular shapes. 
For simplicity, we consider grains to be spherical aggregates with volume $V$. 
Not considering the grains as spherical would require knowing the collision history of the grain, which would greatly increase the complexity of the modelling.

We define the filling factor $\phi$, related to the porosity $\mathfrak{p}$, as the ratio between the volume occupied by the monomers $V_\mathrm{mat}$ and the volume of the aggregate $V$
\begin{equation}\label{Eq:Def_phi}
    \phi = \frac{V_\mathrm{mat}}{V} \quad\text{and}\quad \mathfrak{p} + \phi = 1.
\end{equation}
The mass of an aggregate with mass $m$ and size $s$ can thus be simply computed as
\begin{equation}\label{Eq:Mass_grain}
    m = \rho_\mathrm{s} \phi\frac{4\pi}{3} s^3.
\end{equation}

Two energies can be associated with an aggregate. The first is the kinetic energy upon impact when two identical 
grains with mass $m$ collide with a relative velocity $v_\mathrm{rel}$. In the referential of the centre of mass, the kinetic energy is expressed as
\begin{equation}\label{Eq:Kinetic_energy}
   E_\mathrm{kin}
    =\frac{m}{4}\,v_{\mathrm{rel}}^2,
\end{equation}
where the factor of 1/4 arises from the reduced mass for two grains of identical mass.

The second energy is the rolling energy $E_\mathrm{roll}$, which corresponds to the energy required to rotate a monomer by 
90$^\circ$ around a connection point. The ability of monomers to reorganize leads to internal rearrangement of the 
aggregate \citep{dominik_physics_1997}
\begin{equation}\label{Eq:Rolling_energy}
   E_\mathrm{roll} = 6\pi^2 \gamma_\mathrm{s} a_0 \xi_\mathrm{crit},
\end{equation}
where $\gamma_\mathrm{s}$ is the surface energy (energy per unit area) of a monomer, $\xi_\mathrm{crit}$ the critical rolling displacement and $a_0$ the monomer size. 
The value of $\xi_{\mathrm{crit}}$ is still poorly constrained. According to \citet{dominik_physics_1997}, the 
critical separation $\delta_\mathrm{c}$ between two monomers before they separate, 
\begin{equation}\label{Eq:delta_c}
    \delta_\mathrm{c} = \left(\frac{27\pi^2}{2}\frac{\gamma_\mathrm{s}^2 a_0}{\mathcal{E}^2}\right)^{\frac{1}{3}},
\end{equation}
is of the same order of magnitude as $\xi_{\mathrm{crit}}$ \citep{chokshi_dust_1993}.
Thus we can rewrite Eq.~(\ref{Eq:Rolling_energy}) as
\begin{equation}
    E_\mathrm{roll} = \left(2916\pi^8\frac{\gamma_\mathrm{s}^5 a_0^4}{\mathcal{E}^2}\right)^\frac{1}{3} \approx 302\left(\frac{\gamma_\mathrm{s}^5 a_0^4}{\mathcal{E}^2}\right)^\frac{1}{3},
\end{equation}
with $\mathcal{E}$ the Young's modulus.
Depending on the ratio between these two energies, two different growth regimes can be distinguished with distinct 
porosity evolution depending on the grain mass $m$ \citep{suyama_geometric_2012, okuzumi_rapid_2012}.

\begin{itemize}
    \item In the `hit \& stick' regime, the grains are small and thus coupled to the gas. Collisions occur at very 
    low relative velocities, with a condition on kinetic energy ($E_\mathrm{kin} < 3b\,E_\mathrm{roll}$, \citealt{suyama_numerical_2008}), 
    where $b$ is a numerical factor equal to 0.15 \citep{okuzumi_rapid_2012}.
    \item As the grains grow, the kinetic energy during impacts increases, becoming much larger than the rolling energy. 
    As a result, energy is dissipated by internal restructuring of the grain structure, leading to compaction. This is referred to as the collisional compression regime.
\end{itemize}

\begin{figure}
    \centering
    \includegraphics[width=\columnwidth,clip]{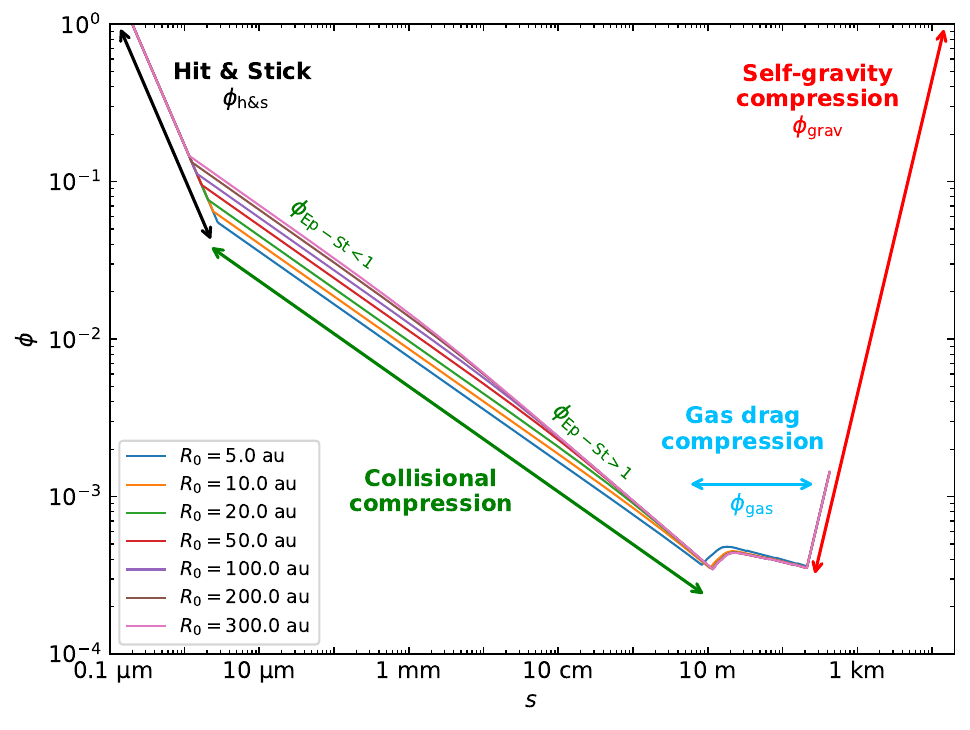}
     \caption{Filling factor $\phi$ as a function of size $s$ for different distances from the star $R_0$ for 
     silicate grains composed of $0.2\ \mu$m monomers in the case of pure growth in our standard disc model (see Sect.~\ref{Ssc:setup}).}
     \label{Fig:fig_croissance_pamdeas}
\end{figure}

We recall in Appendix~\ref{App:Porosity_evol_growth} the equations describing the porosity of dust grains in the different expansion and compression regimes presented in \citet{garcia_evolution_2018} and \citet{garcia_evolution_2020}. Figure~\ref{Fig:fig_croissance_pamdeas} summarises the evolution of the filling factor $\phi$ of a grain, starting as a compact ($\phi=1$) monomer ($s=a_0=0.2\ \mu$m) as its size $s$ increases, for different distances to the star.

\subsection{Bouncing}\label{Ssc:bounce}
In addition to growth, grains can also bounce. We have improved upon the model initially developed by \citet{garcia_evolution_2018}.
\citet{wada_bounce_2011}, \citet{shimaki_experimental_2012}, and \citet{arakawa_2023} have numerically and experimentally shown that 
grains are no longer capable of bouncing when the filling factor drops below $\phi=0.3$. This limit corresponds either to small grains 
composed of a few monomers or to large aggregates spanning several kilometres, when considering only growth -- see Fig.~\ref{Fig:fig_croissance_pamdeas}.
Once below this filling factor, grains can only grow or fragment when colliding.

During bounce, some of the energy is used to bind the grains together, and another part is 
dissipated in a deformation wave \citep{Thornton_Ning_bounce}.
Several velocities can be distinguished. The first is the sticking velocity $v_\mathrm{stick}$, corresponding to the velocity at which grains no longer stick together systematically.
\citet{Johnson_JKR_theory} and \citet{Thornton_Ning_bounce} provide the sticking velocity 
$v_\mathrm{stick}$ for the collision of two identical grains
\begin{equation}\label{Eq:Vstick1}
    v_\mathrm{stick} = 4.23 \left(\frac{\gamma_\mathrm{s}^5 s^4}{m^3 \mathcal{E}^2}\right)^{1/6}.
\end{equation}
\citet{dominik_physics_1997} provide another formula for $v_\mathrm{stick}$ based on the mass and size of monomers. 
Their formula differs by a factor of~2. This factor arises from the medium in which the wave propagates, 
which is the monomer in the case of \citet{dominik_physics_1997}, and a collection of monomers for Eq.~(\ref{Eq:Vstick1}).
However, \citet{shimaki_experimental_2012b,shimaki_experimental_2021} have shown from experiments that the Young's modulus $\mathcal{E}$ of water ice varies with porosity by about an order of magnitude. Its value as a function of $\phi$ is not known for silicates. 
Since $v_\mathrm{stick} \propto \mathcal{E}^{-1/3}$, by using the value of $\mathcal{E}$ for monomers and retaining the factor of 2, a reasonable compromise is achieved. Therefore, we keep
\begin{align}\label{Eq:Vstick2}
    v_\mathrm{stick} &= 8.46 \left(\frac{\gamma_\mathrm{s}^5 s^4}{m^3 \mathcal{E}^2}\right)^{1/6} \\
    &= 2.82 \ \mathrm{mm\,s}^{-1} \left(\frac{\rho_\mathrm{s}}{1000\ \mathrm{kg\,m}^{-3}}\right)^{-1/2} \left(\frac{\gamma_\mathrm{s}}{0.1\ \mathrm{J\,m}^{-2}}\right)^{5/6} \nonumber \\
    &\quad\times \left(\frac{\mathcal{E}}{10\ \mathrm{GPa}}\right)^{-1/3} \left(\frac{s}{1\ \mathrm{mm}}\right)^{-5/6} \phi^{-1/2}, \nonumber
\end{align}
where we have used Eq.~(\ref{Eq:Mass_grain}) to obtain a dependence on size and filling factor. With the material properties we adopt for silicate and water ice grains (see Sect.~\ref{Ssc:setup}), we obtained $v_\mathrm{stick} = 2.89~$mm\,s$^{-1}$ and 5.26~mm\,s$^{-1}$, respectively, for mm-sized grains with $\phi = 0.3$. 
\citet{Weidling_Guttler_dust_sticking} argued that instead of $\gamma_\mathrm{s}$, one should use an effective surface energy, which accounts for porosity and is smaller, together with a smaller value of the Young's modulus for porous aggregates. Both compensate in our expression for $v_\mathrm{stick}$, leading to values that are within one order of magnitude of their experimental data.

However, in their study of collisions of less massive silicate aggregates, \citet{Kothe_2013} found values of $v_\mathrm{stick}$ of several to tens of cm\,s$^{-1}$. Additionally, other studies suggest that, instead of the Young's modulus of the monomers, the compressive strength of the dust aggregates should be used, for which \citet{Blum_2004}, \citet{Guttler_Dynamical_2009} and \citet{Seizinger_2012} found values of $\sim10^3$~Pa for $\phi=0.2$ and $\gtrsim10^4$~Pa for $\phi>0.3$ \citep[see equation~31 in][]{Seizinger_2012}, which increases $v_\mathrm{stick}$ by two orders of magnitude. We thus also consider values of $v_\mathrm{stick}$ 10 to 100 times larger than that given by Eq.~(\ref{Eq:Vstick2}) -- see below.

Two regimes can be distinguished: elastic collision and plastic collision.
The separation between the two regimes is given by the yield velocity $v_\mathrm{yield}$.
\citet{Thornton_Ning_bounce} express $v_\mathrm{yield}$ as a function of the contact radius $a_\mathrm{yield}$ when yield occurs or of the limiting contact pressure $p_\mathrm{yield}$. However, as these material properties are difficult to obtain, and are in particular unknown for water ice, they suggest to turn to values from impact experiments. In theirs, \citet{Weidling_Guttler_dust_sticking} find a value for the yield velocity of SiO$_2$ aggregates about 5 times larger than their sticking velocity. According to \citet{Musiolik_Tieser_ice_grain_coll}, and \citet{Yasui_silicates}, water ice has more favourable sticking properties, and we adopted $v_\mathrm{yield} = 10 v_\mathrm{stick}$. More recently, \citet{kimura_tensile_2020} demonstrated that silicates are more prone to sticking together with a higher surface energy. We thus kept a factor of 10 between the two velocities for silicates as well, and take
\begin{equation}
    v_\mathrm{yield} = 10 v_\mathrm{stick} = 84.6 \left(\frac{\gamma_\mathrm{s}^5 s^4}{m^3 \mathcal{E}^2}\right)^{1/6}.
\end{equation}
Consequently, the coefficient of restitution $e$ can now be defined as follows
\begin{equation}
    e \equiv \frac{v_\mathrm{rel,f}}{v_\mathrm{rel}},
\end{equation}
where $v_\mathrm{rel,f}$ is the relative velocity after bounce. After the collision, a portion of the kinetic energy 
is used for plastic deformation given by $(1-e^2)E_\mathrm{kin}$, and the remaining part serves to separate the grains 
with the remaining kinetic energy $e^2 E_\mathrm{kin}$.
Depending on the relative velocity, the coefficient of restitution can then be expressed 
\citep{Thornton_Ning_bounce, garcia_evolution_2018} as follows
\begin{align}
    v_\mathrm{rel} &\leq v_\mathrm{stick} \quad\quad e=0\\
    v_\mathrm{stick} < v_\mathrm{rel} &\leq v_\mathrm{yield} \quad\quad e=1\\
    v_\mathrm{rel} &\geq v_\mathrm{yield} \quad\quad 
    \label{Eq:Restitcoeff1} 
    e= \left[1.2\sqrt{3}\left(1-\frac{1}{6}\left(\frac{v_\mathrm{yield}}{v_\mathrm{rel}}\right)^2\right)\right. \times \nonumber\\
    &\left.\left(1+2\sqrt{1.2\left( \frac{v_\mathrm{rel}}{v_\mathrm{yield}}\right)^2-0.2}\right)^{-1/2}-\left(\frac{v_\mathrm{stick}}{v_\mathrm{rel}}\right)^2\right]^{1/2}
\end{align}
\begin{figure}
    \centering
    \includegraphics[width=\columnwidth,clip]{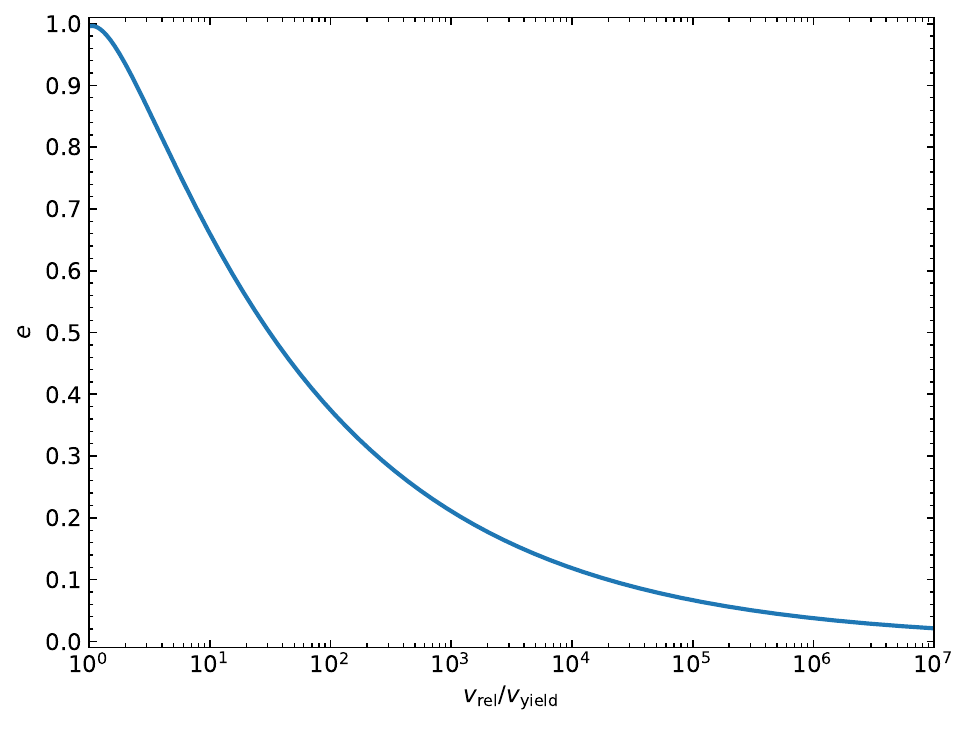}
    \caption{Coefficient of restitution $e$ when $v_\mathrm{rel} \geq v_\mathrm{yield}$, given by Eq.~(\ref{Eq:Restitcoeff1}).}
    \label{Fig:Restitcoeff1}
\end{figure}

When $v_\mathrm{rel} \gg v_\mathrm{yield}$, the coefficient of restitution tends towards zero, as shown in Fig.~\ref{Fig:Restitcoeff1}. This indicates that almost all the energy is used in plastic deformation, and the grain is strongly compacted. Beyond a velocity threshold $v_\mathrm{end}$, the resulting velocity from the remaining kinetic energy $e^2 E_\mathrm{kin}$ becomes so low that after a bounce, the grain is too slow to collide with another grain and can no longer stick. To be consistent with experiments from \citet{shimaki_experimental_2012b}, we set $v_\mathrm{end} = 0.03 v_\mathrm{rel}$. Solving Eq.~(\ref{Eq:Restitcoeff1}) for $e=0.03$ with $v_\mathrm{yield} = 10 v_\mathrm{stick}$ allows to express $v_\mathrm{end}$ as a function of $v_\mathrm{stick}$, so that
\begin{equation}
    v_\mathrm{end} \approx 2\times 10^7 v_\mathrm{stick} \approx 2 \times 10^8 \left(\frac{\gamma_\mathrm{s}^5 s^4}{m^3 \mathcal{E}^2}\right)^{1/6}.
\end{equation}
Typical values of $v_\mathrm{end}$ are of the order of $10-100$~km\,s$^{-1}$, much larger than fragmentation thresholds.

\begin{figure}
   \centering
   \includegraphics[width=\columnwidth,clip]{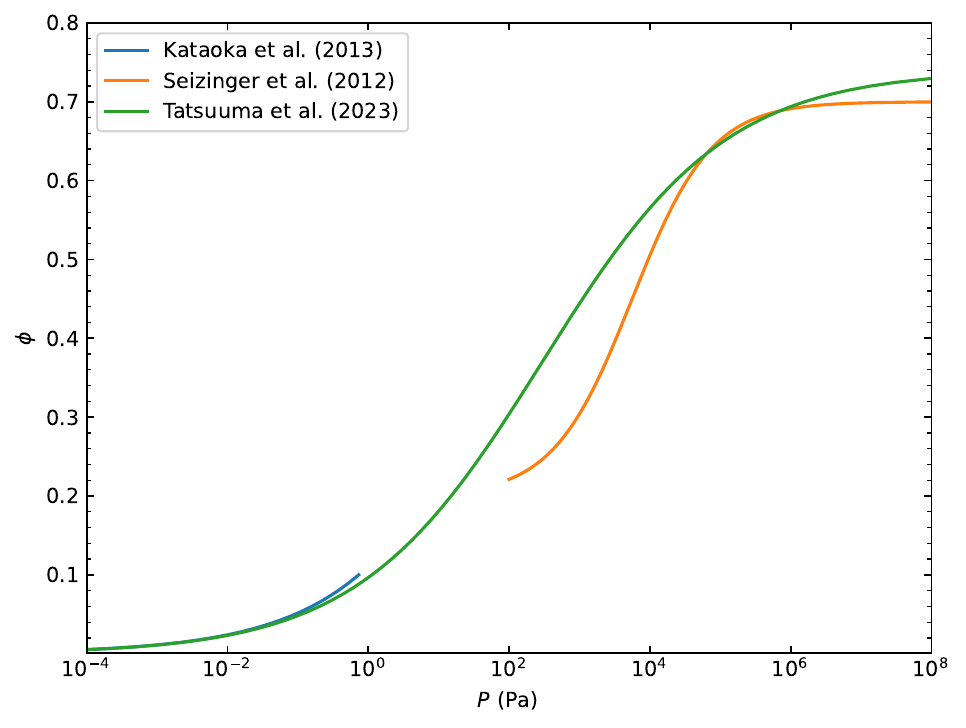}
   \caption{Model comparison for compression pressure. \citet{Tatsuuma_2023} connect the regime with small $\phi$ given by \citet{kataoka_static_comp_2013} to the large $\phi$ with the model of \citet{guttler_outcome_2010} and \citet{Seizinger_2012}.}
   \label{Fig:fig_compression_pressure}
\end{figure}

\subsubsection*{Sticking probability and filling factor}
Empirically, \citet{Weidling_Guttler_dust_sticking} determined the probability that a grain sticks, or not.
As previously mentioned, when $v_\mathrm{stick} < v_\mathrm{rel} < v_\mathrm{end}$, an aggregate can bounce instead of stick. Naturally, the closer $v_\mathrm{rel}$ is to $v_\mathrm{stick}$, the higher the chances of sticking, and conversely for $v_\mathrm{end}$.
Thus, \citet{garcia_evolution_2018}, inspired by \citet{Weidling_Guttler_dust_sticking}, defined the probability $\mathcal{P}$ of sticking for a grain based on its velocity according to
\begin{equation}
    \mathcal{P} = \left\{ 
    \begin{array}{l l}
    \displaystyle  1, & \; v_\mathrm{rel}< v_\mathrm{stick},\\
    \\
    \displaystyle  \frac{\log{\left(v_\mathrm{rel}\right)} - \log{\left(v_\mathrm{end}\right)}}{\log{\left(v_\mathrm{stick}\right)} - \log{\left(v_\mathrm{end}\right)}}, & \; v_\mathrm{stick} \leq v_\mathrm{rel} < v_\mathrm{end},\\ 
    \\
    \displaystyle 0. & \; v_\mathrm{rel}\geq v_\mathrm{end},
    \end{array} \right.
\end{equation}
Depending on the sticking probability, the value of the growth rate will be different. When the grain does not stick, there is no contribution to the growth rate.
Thus, it can be expressed as
\begin{equation}
    \left(\frac{\mathrm{d} m}{\mathrm{d} t}\right)_{\mathrm{bounce}}= \mathcal{P}\left(\frac{\mathrm{d} m}{\mathrm{d} t}\right)_{\mathrm{grow}}.
\end{equation}
However, when considering bouncing, a problem arises. It is impossible to write the filling factor of a grain in terms 
of its mass when the deformation is plastic. \citet{garcia_evolution_2018} provides a formula for the final filling factor during bounce based on the initial filling factor and an arbitrary time step $t$
\begin{equation}\label{Eq:phi_bounce}
    \phi_\mathrm{bounce} \approx \phi_\mathrm{i} \left(1-\frac{\Delta V}{V_\mathrm{i}}\right)^{-n},
\end{equation}
where $n=t/\tau_\mathrm{coll}$ and $\Delta V$ is the volume change during compression due to plastic deformation.
\citet{shimaki_experimental_2012} provided an expression to compute $\Delta V$, which is proportional to the kinetic energy
\begin{equation}\label{Eq:DeltaV_bounce}
    \Delta V = \frac{(1-e^2)E_\mathrm{kin}}{2\varUpsilon_\mathrm{d} },
\end{equation}
where $\varUpsilon_\mathrm{d}=\varUpsilon_0 \phi^\mathcal{N}$ is the dynamic compression resistance, and $\varUpsilon_0$ is 
the reference dynamic compression resistance. \citet{mellor1974review} give $\varUpsilon_0=9.8$ MPa and $\mathcal{N}=4$ for water ice. 
However, $\varUpsilon$, which depends on porosity, is rarely studied because it is difficult to determine, especially for highly porous materials.
Therefore, this formula cannot be used directly as $\varUpsilon$ remains unknown for most materials that one might want to simulate, such as silicates.
However, for numerous bounces, the process can be approximated as static compression. \citet{kataoka_static_comp_2013} provides a formula for 
static compression pressure for highly porous grains only, as given in Eq.~(\ref{Eq:Pcomp}), and \citet{Guttler_Dynamical_2009,guttler_outcome_2010} and \citet{Seizinger_2012} provide an expression for values of $\phi > 0.2$.
Very recently, \citet{Tatsuuma_2023} have developed a formula for compression pressure that depends only on $E_\mathrm{roll}$ and $\phi$ which is valid for both small and large $\phi$, shown on Fig.~\ref{Fig:fig_compression_pressure},
\begin{equation}\label{Eq:P_bounce_Tat}
    P_\mathrm{bounce} = \left(\frac{E_\mathrm{roll}}{a_0^3}\right)\left( \frac{1}{\phi} - \frac{1}{0.74}\right)^{-3}.
\end{equation}
The value $\phi =0.74$ corresponds to the maximum possible filling factor, which is that of either a cubic or a hexagonal close packed arrangement of equal spheres.

\begin{figure}
   \centering
   \includegraphics[width=\columnwidth,clip]{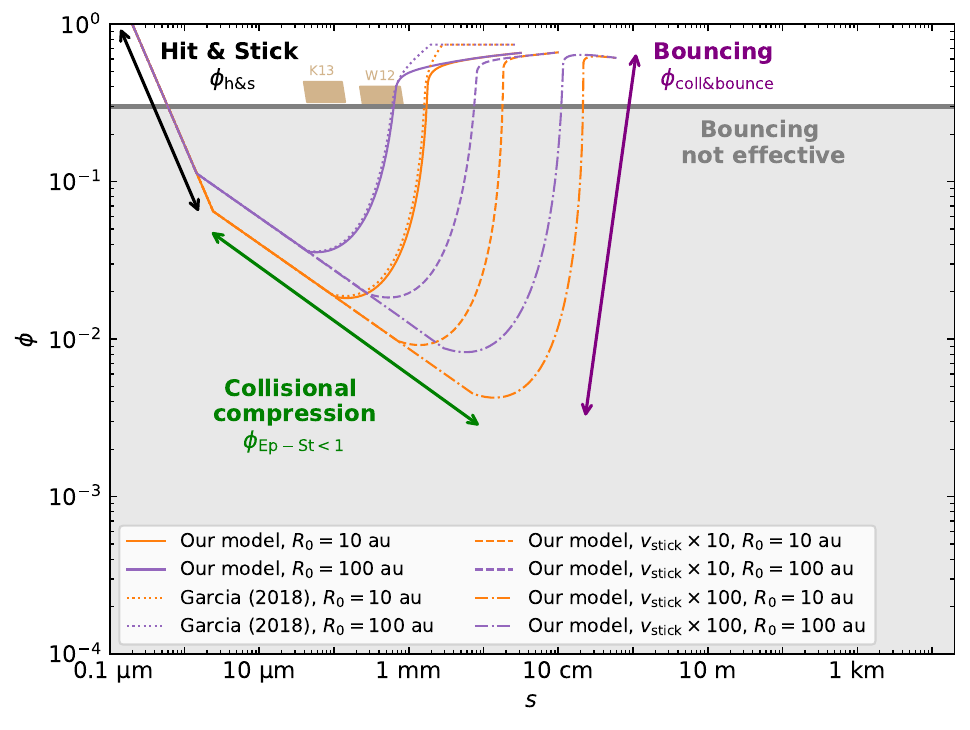}
   \caption{Growth and bouncing of a water ice grain with $a_0=0.2\ \mu$m in our standard disc model (see Sect.~\ref{Ssc:setup}) for two different distances. 
   Our model (solid lines) uses Eq.~(\ref{Eq:P_bounce_Tat}), while the model from \citet[dotted lines]{garcia_evolution_2018} uses the $\varUpsilon_d$ given by \citet{shimaki_experimental_2012}.
   The dashed and dot-dashed lines show the impact of multiplying $v_\mathrm{stick}$ from Eq.~(\ref{Eq:Vstick2}) by 10 or 100. The light tan polygons trace the parameter space probed in experiments by \citet{Weidling_Guttler_dust_sticking} and \citet{Kothe_2013}. The grey shaded area marks the range of filling factors for which bouncing is not effective (see text).}
   \label{Fig:fig_growth+bounce_pamdeas_drift_comparison}
\end{figure}

In order to illustrate what the porosity evolution of a grain experiencing only growth and bouncing would be, we consider for the time being all values of the filling factor before restricting it again to values above $\phi=0.3$. Figure~\ref{Fig:fig_growth+bounce_pamdeas_drift_comparison} allows for the comparison of the bounce model from \citet{garcia_evolution_2018} with the equation involving $\varUpsilon$, 
which is valid only for water ice (or a mixture of ice and silicates), and our model where $P_\mathrm{bounce}$ replaces $\varUpsilon$ in Eq.~(\ref{Eq:DeltaV_bounce}). 
The difference between both models is observed when the grain is highly compacted, with a filling factor 
$\phi>0.5$. With the model from \citet{garcia_evolution_2018}, the grains continue to be efficiently compacted until they reach the maximum possible filling factor. In our model, as grains become highly compacted, it becomes increasingly difficult to compress them. This is due to the fact that as $\phi$ approaches 1, the value of the compression pressure changes only slightly.
The dashed and dot-dashed curves show that considering values of $v_\mathrm{stick}$ 10 or 100 times larger only delays the onset of bouncing but has very little influence on the final value of the filling factor.
The light tan polygons trace the parameter space explored in the experiments from \citet{Weidling_Guttler_dust_sticking} and \citet{Kothe_2013}, as estimated from the latter's figure~7.

We now relax our working assumption and return to the situation where bouncing is only effective for $\phi\geq0.3$, which excludes the grey-shaded area in Fig.~\ref{Fig:fig_growth+bounce_pamdeas_drift_comparison}. Grains experiencing only growth quickly dive below the threshold and never bounce again. A mechanism able to compact grains, such as presented in Sect.~\ref{Ssc:Compaction during fragmentation}, is thus needed for bouncing to reappear. The sizes at which such compacted grains will cross above the threshold depend mainly on the compacting mechanism. Their subsequent evolution depends very little on the value of $v_\mathrm{stick}$ (see Sect.~\ref{Ssc:Compaction during fragmentation}).

In conclusion, to compute the filling factor due to bouncing, we use, for values larger than 0.3 only,
\begin{equation}\label{Eq:phi_coll_bounce}
    \phi_\mathrm{coll\,\&\,bounce} = \left\{ 
    \begin{array}{l l}
    \displaystyle  \phi_\mathrm{coll}, & \; v_\mathrm{rel}<v_\mathrm{yield},\\
    \\
    \displaystyle  \phi_\mathrm{coll}\mathcal{P} +  (1-\mathcal{P})\phi_\mathrm{bounce}, & \;
     v_\mathrm{yield} \leq v_\mathrm{rel} < v_\mathrm{end},\\ 
    \\
    \displaystyle \phi_\mathrm{bounce}, & \; v_\mathrm{rel}\geq v_\mathrm{end},
    \end{array} \right.
\end{equation}
where $\phi_\mathrm{coll}$ is the filling factor resulting from collisions in one of the regimes described in Appendix~\ref{App:Porosity_evol_growth_coll_comp}, and $\phi_\mathrm{bounce}$ is computed using Eq.~(\ref{Eq:phi_bounce}).
The final filling factor is the largest value between $\phi_\mathrm{gas}$, $\phi_\mathrm{grav}$ and $\phi_\mathrm{coll\,\&\,bounce}$. Our model is an improvement over the model from \citet{garcia_evolution_2018}. It now takes into account the saturation effect of compression for filling factors close to one, and can be used with different materials as long as the surface energy, Young's modulus and monomer size are known.

\subsection{Fragmentation}
\label{Ssc:Fragmentation}

The other natural process that appears in the life of grains is fragmentation. 
When the relative velocity between grains exceeds the fragmentation threshold $v_\mathrm{frag}$, 
then the grain fragments instead of growing \citep{Tanaka_1996}. The kinetic energy upon impact 
is such that the grain's structure cannot absorb it, breaking the bonds between the monomers constituting the aggregate.

A model developed by \citet{kobayashi_fragmentation_2010} and used by \citet{vericel_dust_2021}, allows for a gradual fragmentation depending on the value of $v_\mathrm{rel}$ with respect to $v_\mathrm{frag}$
\begin{equation}\label{Eq:Frag_dmdt_Garcia}
   \left(\frac{\mathrm{d} m}{\mathrm{d} t}\right)_{\mathrm{frag}} = -4 \pi\frac{v_{\mathrm{rel}}^3}{v_{\mathrm{rel}}^2 + v_{\mathrm{frag}}^2} \rho_{\mathrm{d}} s^2.
\end{equation}
In this model, when the relative velocity is close to the threshold, the mass loss is less significant. A fragmenting grain loses half 
of its mass after a collision time $\tau_\mathrm{coll}$ ($v_\mathrm{rel} = v_\mathrm{frag}$) or more ($v_\mathrm{rel} > v_\mathrm{frag}$). 
In the case where $v_\mathrm{rel} \gg v_\mathrm{frag}$, we recover the scenario described by \citet{gonzalez_accumulation_2015}, 
where the entire grain fragments after a collision time, independently of $v_\mathrm{rel}$.

In reality, both experimental studies \citep{blum_experiments_2000,shimaki_experimental_2012,Weidling_Guttler_dust_sticking} 
and numerical experiments \citep{dominik_physics_1997,okuzumi_numerical_2009,okuzumi_rapid_2012,kataoka_static_comp_2013,Krijt_Ormel_frag_comp,Planes_2023} 
show more complex behaviours that are challenging to model. Fragmentation involves a cascade of sizes and depends on arameters such as composition, porosity, shape, impact parameter, mass ratio, impact velocity, etc.
The presented model follows the one introduced by \citet{kobayashi_fragmentation_2010} using a fragmentation definition 
that relies on several approximations to capture the essential physics according to the relative collision velocity.
To determine $v_\mathrm{frag}$, an estimate of the energy required to break the bond between two monomers $E_\mathrm{break}$ is given by \citet{dominik_physics_1997},
\begin{equation}
    E_\mathrm{break} = C_\mathrm{break} F_\mathrm{c} \delta_\mathrm{c},
\end{equation}
where $C_\mathrm{break}$ is a constant that they take equal to 1.8. $F_\mathrm{c}$ is the critical pulling force between two monomers
\begin{equation}
    F_\mathrm{c} = 3\pi \gamma_\mathrm{s} a_0.
\end{equation}
We can then derive the expression for $E_\mathrm{break}$
\begin{equation}\label{Eq:Ebreak}
    E_\mathrm{break} \simeq C_\mathrm{break} 48\left(\frac{\gamma_\mathrm{s}^5 a_0^4}{\mathcal{E}^2}\right)^{\frac{1}{3}}.
\end{equation}

To find the total energy, we just need to know the number of monomers and the number of bonds between monomers
\begin{equation}\label{Eq:Efrag}
    E_\mathrm{frag} = \kappa N_\mathrm{tot} E_\mathrm{break},
\end{equation}
where $N_\mathrm{tot}$ is the total number of monomers, and $\kappa$ is a numerical factor representing the number of bonds 
and depending on the species and porosity. For erosion, $\kappa = 1-2$, and for fragmentation, $\kappa = 3-10$ \citep{blum_experiments_2000,Wada_2007}.
We can then define $v_\mathrm{frag}$ in terms of $E_\mathrm{frag}$ as
\begin{equation}\label{Eq:vfrag}
    v_\mathrm{frag} = 2\sqrt{\frac{E_\mathrm{frag}}{m}}= 2\sqrt{\frac{48\kappa N_\mathrm{tot}C_\mathrm{break}}{m}}\left(\frac{\gamma_\mathrm{s}^5 a_0^4}{\mathcal{E}^2}\right)^{\frac{1}{6}}.
\end{equation}
These equations depend on $\gamma_\mathrm{s}$ and $a_0$. An error in estimating these parameters can thus lead to 
changes in the values of $E_\mathrm{frag}$ and $v_\mathrm{frag}$.

Research on the value of $v_\mathrm{frag}$ according to different types of materials is an active topic, and the community does 
not yet agree on the values. Different studies have yielded a range of values for different materials like silicates and water, 
and there is a complex interplay of factors.
For example, \citet{blum_growth_2008} and \citet{guttler_outcome_2010} deduced $v_{\mathrm{frag,\: Si}}\sim 1\ \msec$ 
for silicates. \citet{wada_collisional_2009,wada_growth_2013} found a similar value for silicates, around 
$v_{\mathrm{frag,\: Si}}\sim 5\ \msec$, as well as that for water, $v_{\mathrm{frag,\: H_2O}}\sim 60-70\ \msec$. 
However, the method used by \citet{yamamoto_examination_2014} yields a different value for water, $v_{\mathrm{frag,\: H_2O}}=56\ \msec$, 
based on the link between surface energy (and hence fragmentation threshold) and material melting temperature. The values for silicates 
differ widely: \citet{yamamoto_examination_2014} found $\gamma_\mathrm{s} = 0.3\ \mathrm{J/m^2}$, while \citet{kimura_tensile_2020} found $\gamma_\mathrm{s} = 0.15\ \mathrm{J/m^2}$. Silicates are stronger than previously thought due to experimental issues with the original estimations of $\gamma_\mathrm{s}$. 
Silicates easily absorb water from the air, which reduces the surface energy of the original aggregate, thereby reducing the adhesion ability of the silicate grains.
In the case of water ice, \citet{shimaki_experimental_2012} measured $v_{\mathrm{frag,\: H_2O}}\sim15\ \msec$, while collision simulations 
gave higher values \citep{wada_collisional_2009}. The same value was computed by \citet{gonzalez_accumulation_2015} using the experimental measurements of the energy required to fragment a unit mass of grains ($\sim 55\ \mathrm{J/kg}$).
For the purpose of comparison with certain previous studies \citep{garcia_evolution_2020,vericel_dust_2021}, we adopt the value of $v_\mathrm{frag,\ H_2O} \approx 15\ \msec$ for water.

\begin{figure*}
    \centering
    \includegraphics[width=0.49\textwidth,clip,trim=.1cm .1cm .1cm .1cm]{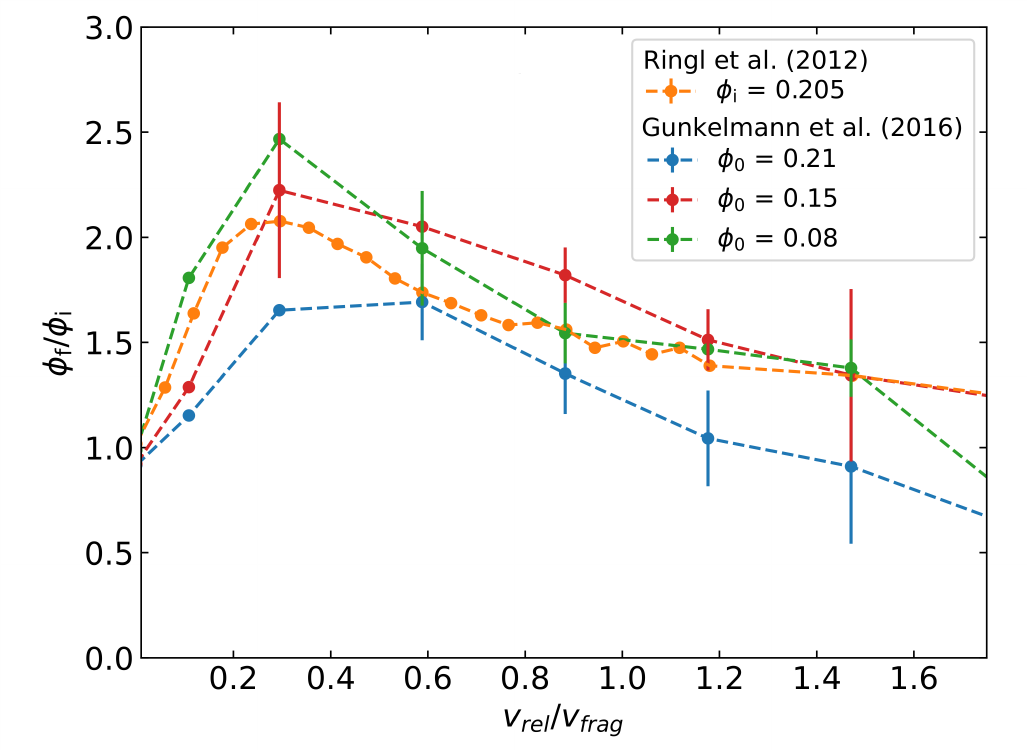}
    \includegraphics[width=0.49\textwidth,clip,trim=.1cm .1cm .1cm .1cm]{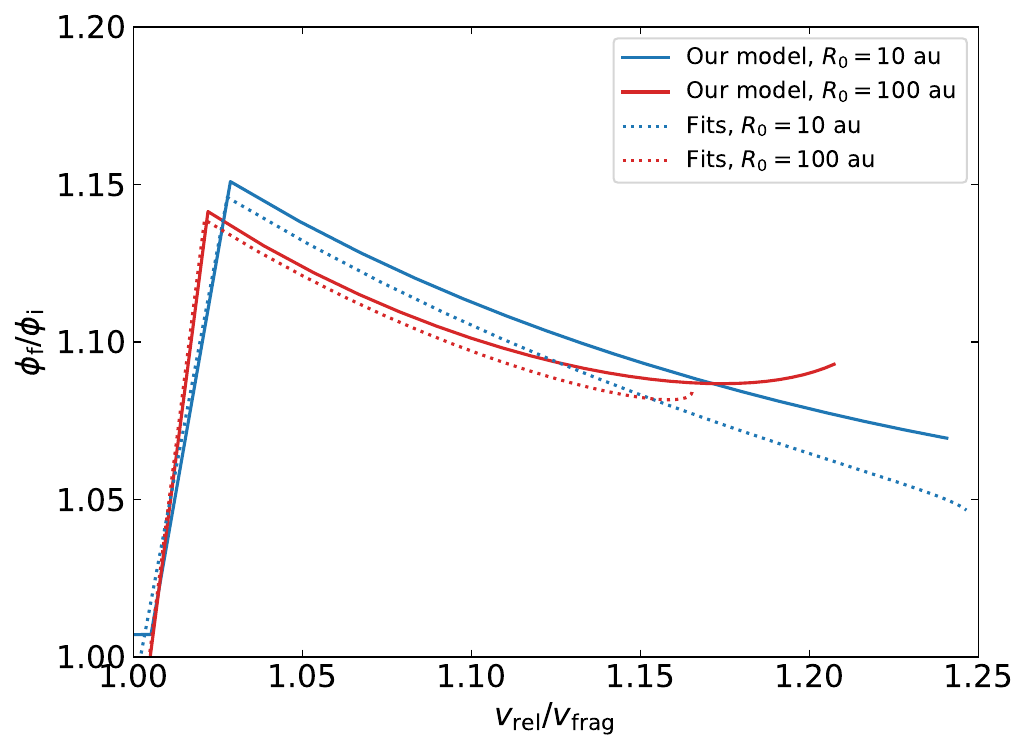}
    \caption{Filling factor $\phi_\mathrm{f}/\phi_\mathrm{i}$ as a function of $v_\mathrm{rel}$.
    \textbf{Left}: Experimental values for different initial filling factors $\phi_\mathrm{i}$ provided by \citet{Ringl_Bringa_2012} in orange and \citet{gunkelmann_influence_2016} in blue, red and green.
    \textbf{Right}: Comparison between our model given by Eq.~(\ref{Eq:phi_frag}), and fits of the curves given by \citet{Ringl_Bringa_2012} and \citet{gunkelmann_influence_2016} in our fragmentation module. 
    }
    \label{Fig:Gunkelmann_Ringl_compaction_frag}
\end{figure*}

However, pure water ice grains are unlikely. Most often, grains are composed of solid materials like silicates, covered by a layer of another more volatile element such as water or organic matter. The fragmentation velocity is primarily determined by the material forming the surface layer, and not the internal material.
Therefore, we choose different thresholds for the silicates, as there are still uncertainties about their values, to mimic the effect of a layer of volatiles: $v_{\mathrm{frag}} = 5$, 10, or $20\ \msec$
for more fragile materials like water or CO, and $v_{\mathrm{frag}} = 40\ \msec$ for modelling pure silicate grains or those surrounded by organic materials.

Contrary to \cite{shimaki_experimental_2012}, who demonstrated that a mixture of silicates and water is more fragile than pure ice, we assume the opposite here, based on measurements conducted in recent years \citep{kimura_tensile_2020,san_sebastian_tensile_2020}. Pure silicate grains are now considered more resistant than water ice grains and we suppose a combination of the two materials is more robust than water ice but less so than silicates. However, this remains to be confirmed experimentally.
Finally, one should note that we assumed surface energies and fragmentation thresholds to be independent of monomer size for the values we have chosen.

\subsection{Compaction during fragmentation}
\label{Ssc:Compaction during fragmentation}

The second improvement we made to the porosity model is to consider grain compaction during fragmentation. 
The main motivation behind this enhancement is to attempt to explain the high filling factor values for intermediate-sized grains (10 $\mu$m-cm) 
found on comets \citep{guttler_synthesis_2019} and inferred from observations through polarimetric measurements 
\citep{Kataoka_polar_2015, Kataoka_polar_2016, Tazaki_kataoka_2019, Kataoka_2019_polar, tazaki_how_2022}. 
Since neither growth, static gas compaction, nor bouncing can account for such high factors ($\phi > 0.1$) as grains grow by 
increasing their porosity, we turned to fragmentation to obtain more compact grains.
\citet{sirono_conditions_2004} showed that the filling factor remains constant during fragmentation. 
For simplicity, this assumption, which considers that the impact energy is used to break bonds between monomers rather than for internal grain restructuring, was adopted by \citet{garcia_evolution_2018}.

However, \citet{ringl_collisions_2012} and \citet{gunkelmann_influence_2016} found that the filling factor after 
fragmentation is approximately 1.5 to 2 times larger than the initial filling factor.
The remaining energy for a grain to undergo compaction can be computed by subtracting the energy $E_\mathrm{frag}$ used to fragment the grain 
from the kinetic energy $E_\mathrm{kin}$ released upon impact. We can rewrite $E_\mathrm{frag}$ as the energy $E_\mathrm{break}$ required to break bonds 
between monomers, multiplied by the number of ejected monomers
\begin{equation}\label{Eq:Ecomp}
    E_\mathrm{comp} = E_\mathrm{kin} - E_\mathrm{frag} = E_\mathrm{kin} - \frac{(2m_\mathrm{i} - m_\mathrm{f})}{m_0}\kappa E_\mathrm{break}.
\end{equation}
Here, $(2m_\mathrm{i} - m_\mathrm{f})/m_0$ corresponds to the number of monomers that were ejected during fragmentation. In the case of fragmentation, 
surface monomers are the first to be ejected since they have fewer bonds \citep{ringl_collisions_2012, gunkelmann_influence_2016}. 
We will therefore consider $\kappa = 3$ defined by Eq.~(\ref{Eq:Efrag}), which has a lower value for fragmentation compared 
to those proposed by \citet{blum_experiments_2000} and \citet{Wada_2007}.
In this sense, we model the fact that ejected grains are closer to the surface, and the remaining energy is used for rearranging 
the internal monomers. A value of $\kappa = 10$ means more energy is used to break bonds, allowing aggregates to remain at 
larger sizes before starting to compact at a constant fragmentation threshold.
We can compute, similarly to bouncing (Eq.~(\ref{Eq:DeltaV_bounce})), the volume change after fragmentation $\Delta V$, which is given by
\begin{equation}
    \Delta V = \frac{E_\mathrm{kin} - (2m_\mathrm{i} - m_\mathrm{f})\kappa E_\mathrm{break}m_0^{-1}}{2 \displaystyle\left(\frac{E_\mathrm{roll}}{a_0^3}\right)\left( \frac{1}{\phi} - \frac{1}{0.74}\right)^{-3}}.
\end{equation}

\begin{figure*}
    \centering
    \includegraphics[width=0.49\textwidth,clip,trim=.1cm .1cm .1cm .1cm]{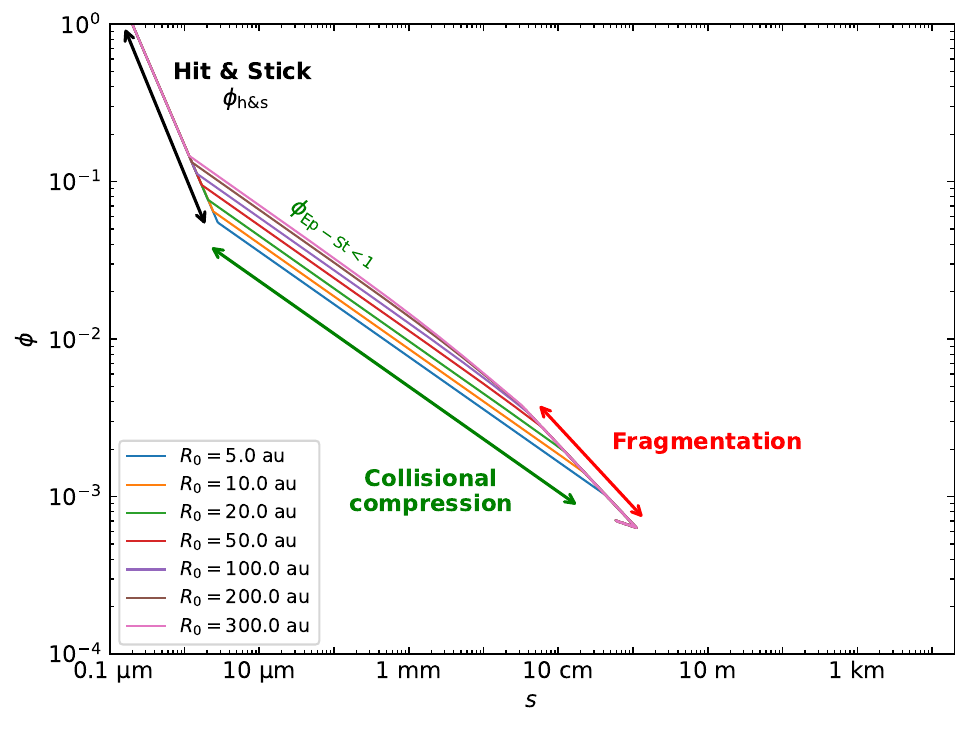}
    \includegraphics[width=0.49\textwidth,clip]{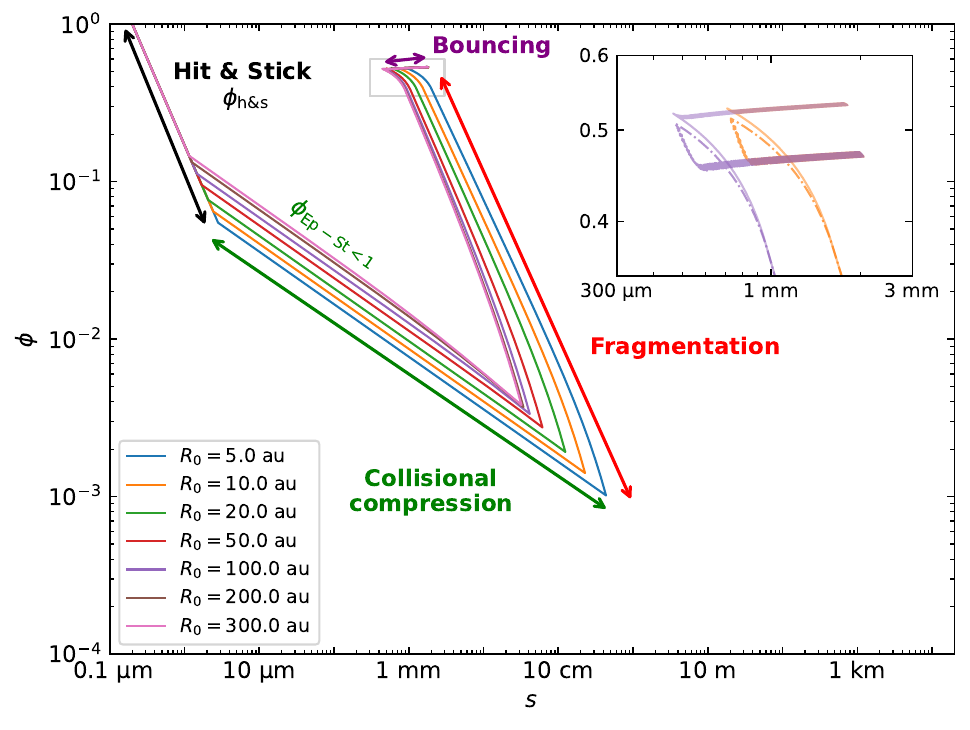}
    \caption{Filling factor $\phi$ as a function of size $s$ for different distances to the star $r$ for 
    silicate grains composed of $0.2\ \mu$m monomers in our standard disc model with growth and fragmentation.
    \textbf{Left}: Without compaction.
    \textbf{Right}: With compaction. The inset is a zoom of the light grey box showing only tracks for $R_0=10$ and 100~au. The dash-dotted lines show the impact of multiplying $v_\mathrm{stick}$ from Eq.~(\ref{Eq:Vstick2}) by 100.}
     \label{Fig:fig_croissance_fragcomp_rebond_pamdeas_drift}
\end{figure*}

Furthermore, we fitted the data obtained by \citet{ringl_collisions_2012} and \citet{gunkelmann_influence_2016}, see the left panel of Fig.~\ref{Fig:Gunkelmann_Ringl_compaction_frag}, which shows the ratio $\phi_\mathrm{f}/\phi_\mathrm{i}$ of the final filling factor to the initial one as a function of the relative velocity $v_\mathrm{rel}$ for three values of $\phi_\mathrm{i}$.
It should be noted that \citet{gunkelmann_influence_2016} define $v_\mathrm{frag}$ as the point when a grain starts to lose a monomer, which is equal to 0.17 $\msec$, and differs from our definition as the point where the grain loses half of its mass.
We deduce that the ratio $\phi_\mathrm{f}/\phi_\mathrm{i}$ varies as $\exp{\left(1-\left(v_\mathrm{rel}/v_\mathrm{frag}\right)^2\right)}$.
In the right panel of Fig.~\ref{Fig:Gunkelmann_Ringl_compaction_frag}, we present a comparison between our model and the fits to data from \citet{ringl_collisions_2012} and \citet{gunkelmann_influence_2016} for two distances to the star, 10 and 100 au. Our model shows good agreement with the fits. Divergence between the model and fits occurs when $v_\mathrm{rel}/v_\mathrm{frag}$ exceeds 1.15 to 1.2. However, grains never reach these values as they fragment before reaching such velocities. 
One should note the axes are different. On the left panel, the initial filling factors are constant for all values of $v_\mathrm{rel}/v_\mathrm{frag}$. On the contrary, on the right panel, the fit extracted from \citet{ringl_collisions_2012} and \citet{gunkelmann_influence_2016} has been implemented in our growth and fragmentation model and compared to our compaction model. In this case, the value of $\phi_0$ varies after each collision.
The filling factor after fragmentation is computed in the same way as bouncing, using Eq.~(\ref{Eq:phi_bounce}).
\begin{equation}\label{Eq:phi_frag}
    \phi_\mathrm{frag-comp} \approx \phi_\mathrm{i} \left(1-\exp\left(1-\left(\frac{v_\mathrm{rel}}{v_\mathrm{frag}}\right)^2\right)\frac{\Delta V}{V_\mathrm{i}}\right)^{-n},
\end{equation}
where $n$ represents the number of collisions during a time step.
The effect of compaction can be seen in Fig.~\ref{Fig:fig_croissance_fragcomp_rebond_pamdeas_drift}, produced with \textsc{Pamdeas} with growth, bouncing, fragmentation and radial drift (see Sect.~\ref{Ssc:Pamdeas}). In the left panel, where compaction is not taken into account, grains reach a growth-fragmentation equilibrium and stay in the same range of size and filling factor. $\phi$ remains below 0.3, bouncing thus never occurs. In the right panel, with compaction, the filling factor of fragmenting grains increases while their size decreases. Because $\phi$ reaches values larger than 0.3, bouncing then appears and must be considered. It prevents grains from growing significantly when compacted. The inset shows that using a value of $v_\mathrm{stick}$ 100 times larger than that given by Eq.~(\ref{Eq:Vstick2}) -- see Sect.~\ref{Ssc:bounce} -- has a very limited impact and only slightly lowers the final filling factor.

\subsection{Rotational disruption}
\label{Ssc:Disruption}

Rotational disruption was identified as a possible growth barrier in protoplanetary discs by \citet{tatsuuma_rotational_2021}: highly porous grains can be disrupted by the gas-flow torque when the tensile stress from the centrifugal force exceeds their tensile strength. \citet{Michoulier_Gonzalez_Disruption} studied the effects of rotational disruption in 1D simulations, in this work we implemented their equations in \textsc{Phantom}.

\subsection{Summary}

The algorithm to compute the final filling factor can be summarised as follows:
\begin{algorithmic}
\State Compute $\phi_\mathrm{coll}$, $\phi_\mathrm{gas}$, $\phi_\mathrm{grav}$
\State Compute $\phi_\mathrm{min}= \max(\phi_\mathrm{coll}$, $\phi_\mathrm{gas}$, $\phi_\mathrm{grav})$
\If{$v_\mathrm{rel} < v_\mathrm{frag}$}
\State Compute $\phi_\mathrm{grow}$
    \If{(Grains can bounce)}
    \State Compute $\phi_\mathrm{coll\,\&\,bounce}$
    \State Compute $\phi_\mathrm{f}=\max(\phi_\mathrm{coll\,\&\,bounce}$, $\phi_\mathrm{min})$
    \Else
    \State Compute $\phi_\mathrm{f}=\max(\phi_\mathrm{grow}$, $\phi_\mathrm{min})$
    \EndIf
\Else
    \If{(Fragmentation with compaction)}
        \State Compute $\phi_\mathrm{frag-comp}$
        \State Compute $\phi_\mathrm{f}=\max(\phi_\mathrm{frag-comp}$, $\phi_\mathrm{min})$
    \Else
         \State Compute $\phi_\mathrm{f}=\max(\phi_\mathrm{i},\phi_\mathrm{min})$
    \EndIf  
\EndIf 
\State $\phi_\mathrm{final}=\min(\phi_\mathrm{f},0.74)$
\end{algorithmic}

\section{Numerical simulations}\label{Sc:Simulations}

\subsection{The 1D code \textsc{Pamdeas}}\label{Ssc:Pamdeas}

\begin{figure*}
  \centering
  \includegraphics[width=0.49\textwidth,clip]{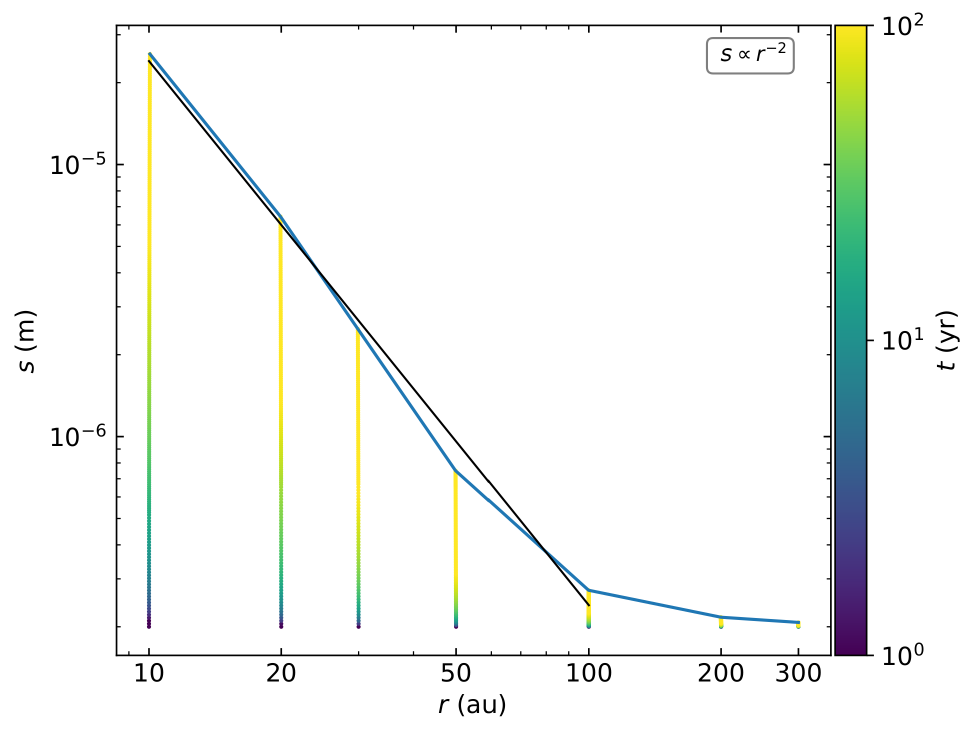}
  \includegraphics[width=0.49\textwidth,clip]{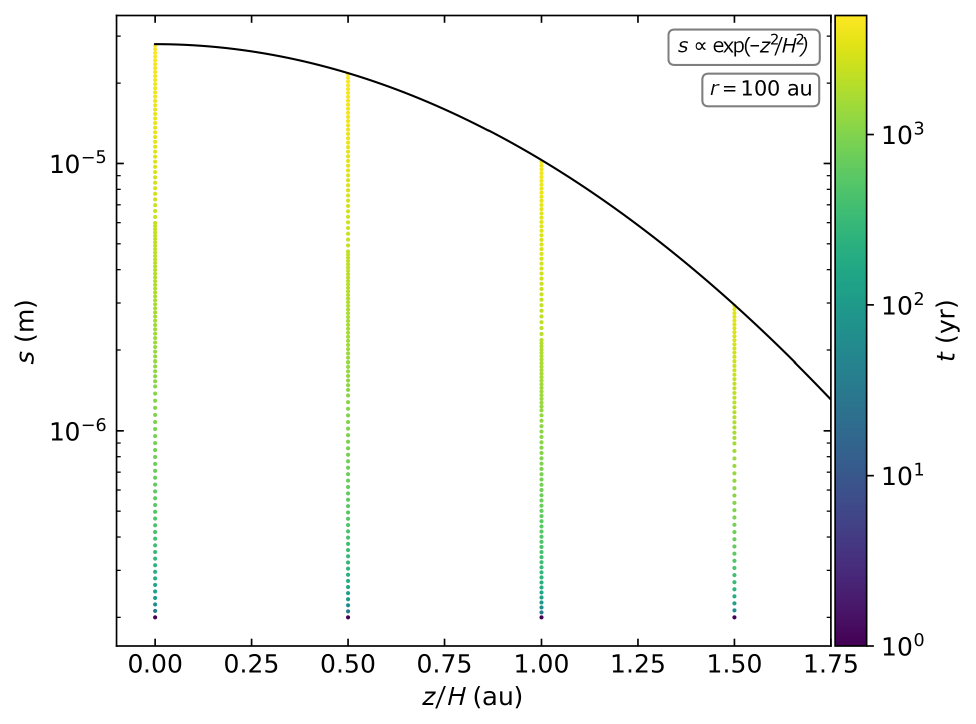}
   \caption{Radial (left panel) and vertical (right panel) profiles of size $s$ obtained with \textsc{Pamdeas} using our standard disc model after 100~yr. Grains grow from the monomer size at different initial positions. Colour indicates time $t$.}
   \label{Fig:size_distrib}
\end{figure*}

\textsc{Pamdeas} (Porous Aggregate Model and Dust Evolution in protoplAnetary discS) is a one-dimensional code, presented in \citet{Michoulier_Gonzalez_Disruption}, to study the evolution of porous grains  within protoplanetary discs considering different physical processes, such as growth, fragmentation, radial drift or rotational disruption. We added new physics such as aeolian erosion \citep{Rozner_2020_Erosiona,Grishin_2020_Erosionb,Michoulier_erosion}, bouncing and compaction during fragmentation. This code allows to follow the evolution of a given number of particles from a set of initial conditions in a static, vertically isothermal, non self-gravitating gas disc. The gas surface density and temperature profiles are given by power-laws of indices $p$ and $q$, respectively
\begin{equation}\label{Eq:Sigma_power-law}
    \Sigma_\mathrm{g}(r)=\Sigma_\mathrm{g,\ 0}(r/R_0)^{-p}
\end{equation}
and 
\begin{equation}\label{Eq:T_power-law}
    T_\mathrm{g}(r)=T_\mathrm{g,\ 0}(r/R_0)^{-q},
\end{equation}
where $R_0$ is a reference radius. The disc extends from $R_{\mathrm{in}}$ to $R_{\mathrm{out}}$ and has a total mass $M_\mathrm{disc}$. As the gas disc structure is held fixed, the dust-to-gas ratio $\varepsilon$ is kept constant and the feedback of dust on gas is neglected. Each grain evolves separately, starting from an initial size equal to the monomer size, and the code tracks its full evolution. Despite its limitations, simulations with \textsc{Pamdeas} are useful to understand the different stages experienced by a single grain.

\subsection{The 3D SPH code \textsc{Phantom}}\label{Ssc:Phantom}

\textsc{Phantom} \citep{price_phantom_2018} is a 3D Smoothed Particle Hydrodynamics (SPH; \citealt{lucy_1977,gingold_monaghan_1977}) code for hydrodynamics and magnetohydrodynamics, designed to be efficient. It is public and widely used. In this paper, we introduce a new module, in which we implemented the algorithms to treat dust porosity with all the physics presented in Sect.~\ref{Ssc:Porosity Model}--\ref{Ssc:Disruption}, both in the `dust-as-mixture' (also known as `one-fluid') and `dust-as-particles' (or `two-fluid') formalisms. This development is tightly coupled to the dust growth module implemented and described in detail by \citet{vericel_dust_2021}. \textsc{Phantom}, which takes into account collective effects and dust vertical settling, and simulates the coupled evolution of gas and dust due to aerodynamic drag (including the back-reaction of dust on gas), produces more realistic dust distributions that can be compared to observations of discs.

We first setup a gas disc with an exponentially decreasing surface density profile
\begin{equation}\label{Eq:Sigma_power-law-exp}
    \Sigma(r) = \Sigma_0 \left(1-\sqrt{\frac{R_\mathrm{in}}{r}}\right) \left(\frac{r}{R_\mathrm{c}}\right)^{-p}\exp{\left[-\left(r/R_\mathrm{c}\right)^{2-p}\right]},
\end{equation} 
where $R_\mathrm{c}$ is the cut-off radius. 
With the exponential tapering, the outer density profile is smooth, rather than being sharply truncated at $R_{\mathrm{out}}$, allowing better control over how relaxation behaves in the outer region. Furthermore, using a pseudo-relaxed disc as the initial state makes the computation faster as particles do not have to significantly move radially. Simulating the innermost region with sufficient resolution is computationally expensive because it leads to a large dynamic time ratio between the inner and outer edges. As a result, we set the accretion radius, which represents the radius within which a particle is considered accreted, to $R_{\mathrm{in}}$. We also remove any particle moving further than 1000~au away to prevent completely isolated particles.
The gas is locally isothermal, with a temperature profile set by Eq.~(\ref{Eq:T_power-law}), and non self-gravitating.

Similarly to \citet{Price_1fluid_2015}, and in contrast to 
\citet{vericel_dust_2021}, we first perform a simulation solely with gas to avoid any spurious behaviour of the dust grains during gas relaxation. We increase the number of particles and the disc mass by about 20\% with respect to the target values to compensate for the mass loss due to particle accretion onto the star during relaxation. When the density profile of the outer gas region stabilizes, typically after 7 to 8 orbits at the outer edge, the disc is fully relaxed.

We then add the dust phase with the same spatial distribution as the relaxed gas, with an initially uniform dust-to-gas ratio $\varepsilon_0$. Unlike in \textsc{Pamdeas}, we do not take the monomer size as the initial grain size. Indeed, if the simulation was initialized with a single, user-defined, value of the initial size for all grains, the filling factor computed from $s$ with Eq.~(\ref{Eq:phi_hs_size}) and the mass computed from Eq.~(\ref{Eq:Mass_grain}) would result in less porous, more massive grains in the upper disc layers. This is unphysical. Furthermore grains in the inner region are expected to have grown during the early stages of star formation, as shown in \citet{Bate_2022} and \citet{Lebreuilly_2023}. We thus use an initial state where each SPH particle has a size depending on its location ($r$,$z$), following fits to a \textsc{Pamdeas} simulation of the evolution of grains from monomers for 100~yr. Figure~\ref{Fig:size_distrib} shows the resulting radial and vertical size profiles. The left panel shows that the midplane radial size distribution can be fitted by a power-law
\begin{equation}\label{Eq:size_profil}
    s(r)=s_0 (r/R_0)^{-2},
\end{equation}
and in the right panel, the vertical dependence of the final size is $\propto\exp{(-z^2/H^2)}$, where $H$ is the disc scale height, is obtained, here at 100 au. After the initialization, the evolution variable is the grain mass, and the mass growth rate and filling factor are computed using the equations in Sect.~\ref{Sc:Methods}.

\begin{table*}
	\centering
	\caption{List of simulations carried out with both \textsc{Pamdeas} and \textsc{Phantom}. `G' stands for growth, `F' for fragmentation, `B' for bounce, `c' for compaction and `D' for rotational disruption. The term `comp' in the column for the monomer size $a_0$ denotes compact grains, with porosity evolution not taken into account. 
   *This simulation includes a snow line, indicated by `S', with an inner threshold of $20\ \mathrm{\msec}$ and an outer threshold of $5\ \mathrm{\msec}$ 
   to model the CO snow line ($T=20$ K, approximately $\sim$ 100 au).
   `D' simulations were run only with \textsc{Phantom}.}
	\label{tab:simulations_names}
	\begin{tabular}{*{6}{c}}
		\hline\hline
		Name & species & bouncing $\&$ compaction & disruption & $a_0$ ($\mu$m) & $v_\mathrm{frag}$ ($\msec$)\\
		\hline
    GF-Si-comp-Vf10    & Si  & no  & no  & comp & 10 \\
    GF-Si-comp-Vf20    & Si  & no  & no  & comp & 20 \\
    GF-Si-comp-Vf40    & Si  & no  & no  & comp & 40 \\

    GF-Si-a02-Vf10     & Si  & no  & no  & 0.2  & 10 \\
    GF-Si-a02-Vf20     & Si  & no  & no  & 0.2  & 20 \\
    GF-Si-a02-Vf40     & Si  & no  & no  & 0.2  & 40 \\
  
    GBFc-Si-a02-Vf10   & Si  & yes & no  & 0.2  & 10 \\
    GBFc-Si-a02-Vf20   & Si  & yes & no  & 0.2  & 20 \\
    GBFc-Si-a02-Vf40   & Si  & yes & no  & 0.2  & 40 \\

    GBFcS-Si-a02-Vf205*& Si  & yes & no  & 0.2  & 20 \\

    GF-H2O-comp-Vf15   & H2O & no  & no  & comp & 15 \\
    GF-H2O-a02-Vf15    & H2O & no  & no  & 0.2  & 15 \\
    GBFc-H2O-a02-Vf15  & H2O & yes & no  & 0.2  & 15 \\

    GBFcD-Si-a02-Vf10  & Si  & yes & yes & 0.2  & 10 \\
    GBFcD-Si-a02-Vf20  & Si  & yes & yes & 0.2  & 20 \\
    GBFcD-H20-a02-Vf15 & H2O & yes & yes & 0.2  & 15 \\
	\hline
	\end{tabular}
\end{table*}

In this paper, the gas and dust disc is evolved as a single set of SPH particles, using the dust-as-mixture algorithms of \citet{Price_1fluid_2015} and \citet{Ballabio_2018}, based on \citet{Laibe_2014a}.

\subsection{Setup}\label{Ssc:setup}

We have chosen to use a disc model which represents an `average disc' \citep{williams_parametric_2014}. The mass of the star is fixed at $M_\mathrm{star} = 1\ \msol$, and the mass of the disc is $M_\mathrm{disc} = 0.01\ \msol$. The temperature is set by the choice of the aspect ratio $(H/R)_0=0.0895$ at $R_0=100$~au, with $q=0.5$.

For \textsc{Pamdeas}, we use $R_{\mathrm{in}} = 1$ au, $R_{\mathrm{out}} = 300$ au and $p=1$. All grains evolve from an initial size equal to the monomer size. We run the simulations up to $t=1$~Myr.

For \textsc{Phantom}, we take $R_{\mathrm{in}} = 10$ au, $R_{\mathrm{out}} = 400$ au and $p=0.75$. This value of $p$, combined with the exponential tapering, leads to a profile whose slope, after relaxation, is similar to that of \textsc{Pamdeas} between 10 and 300~au. We set the number of particles to 1.2 million, the turbulent viscosity parameter \citep{shakura_black_1973} to $\alpha = 5\times 10^{-3}$ (by setting $\alpha_{\rm AV} = 0.1658$) and the initial dust-to-gas ratio $\varepsilon_0$ to a typical value of 1\%.  We evolve the simulations for 300,000~yr.

For both codes, the size of monomers is set to $a_0 = 0.2\ \mu$m, in agreement with recent observations. 
Indeed, \citet{tazaki_how_2022} and \citet{Tazaki_monomers_2023} have shown that to accurately reproduce the polarization degree with respect 
to the scattering angle and the SED, the small grains present on the surface of the IM Lupi disc are most likely fractal aggregates with 0.2 $\mu$m monomers. \citet{Verrios_2022} reached the same conclusion with respect to the dust settling in IM Lupi, requiring porosities of $\phi \lesssim 0.1$ to match the observations. We have chosen to use two different species commonly found in discs: water ice and silicates.

For silicate grains, we have chosen an intrinsic density of $\rho_\mathrm{s} = 2\,700\ \kgmcube$ and a surface energy $\gamma_\mathrm{s} = 0.2\ \jmsqare$, 
in agreement with \citet{yamamoto_examination_2014}, who estimate $\gamma_\mathrm{s} = 0.3\ \jmsqare$. This value is of the same order of magnitude as $\gamma_\mathrm{s} = 0.15\ \jmsqare$, found by \citet{kimura_cohesion_2015} and \citet{kimura_tensile_2020}, through experimental measurements 
with sicastar$^\circledR$ aggregates (micromod Partikeltechnologie GmbH). The Young's modulus is $\mathcal{E} = 72$ GPa \citep{yamamoto_examination_2014}, 
which implies, assuming that the critical separation between two monomers $\delta_\mathrm{c}$ before their separation is of the same order of magnitude as $\xi_\mathrm{crit}$ \citep{chokshi_dust_1993}, a value of
$\xi_\mathrm{crit} \approx 6$~\AA.

For pure water ice, the intrinsic density of monomers is $\rho_\mathrm{s} = 1\,000\ \kgmcube$, with a surface energy of $\gamma_\mathrm{s} = 0.1\ \jmsqare$. The critical separation is still debated, as are many other properties of dust in protoplanetary discs. We have chosen a Young's modulus of $\mathcal{E} = 9.4$ GPa, following \citet{yamamoto_examination_2014}, which results in a value of $\xi_{\mathrm{crit}} \sim 10$ \AA, 
close to $\xi_{\mathrm{crit}} \sim 8$ \AA\ used by \citet{wada_bounce_2011} and \citet{tatsuuma_rotational_2021}.

The different values of the fragmentation thresholds for both species are given in Sect.~\ref{Ssc:Fragmentation}. In our simulations, silicates serve as the reference material, and $v_{\mathrm{frag}} = 20\ \msec$ is the reference threshold. 
We have conducted numerous simulations with different parameters. 
For easier reading, each simulation is assigned a name, as listed in Table~\ref{tab:simulations_names}.

\section{Results}\label{Sc:Results}

\subsection{Effect of porosity on grain growth}\label{Ssc:porosity_effect}
In this section, we compare the evolution of compact (i.e., $\phi=1$) and porous dust, including only growth and fragmentation for now.

\subsubsection{1D Simulations with \textsc{Pamdeas}}\label{Ssc:porosity_effect_1D_simulations}

\begin{figure}
  \centering
  \includegraphics[width=\columnwidth,clip]{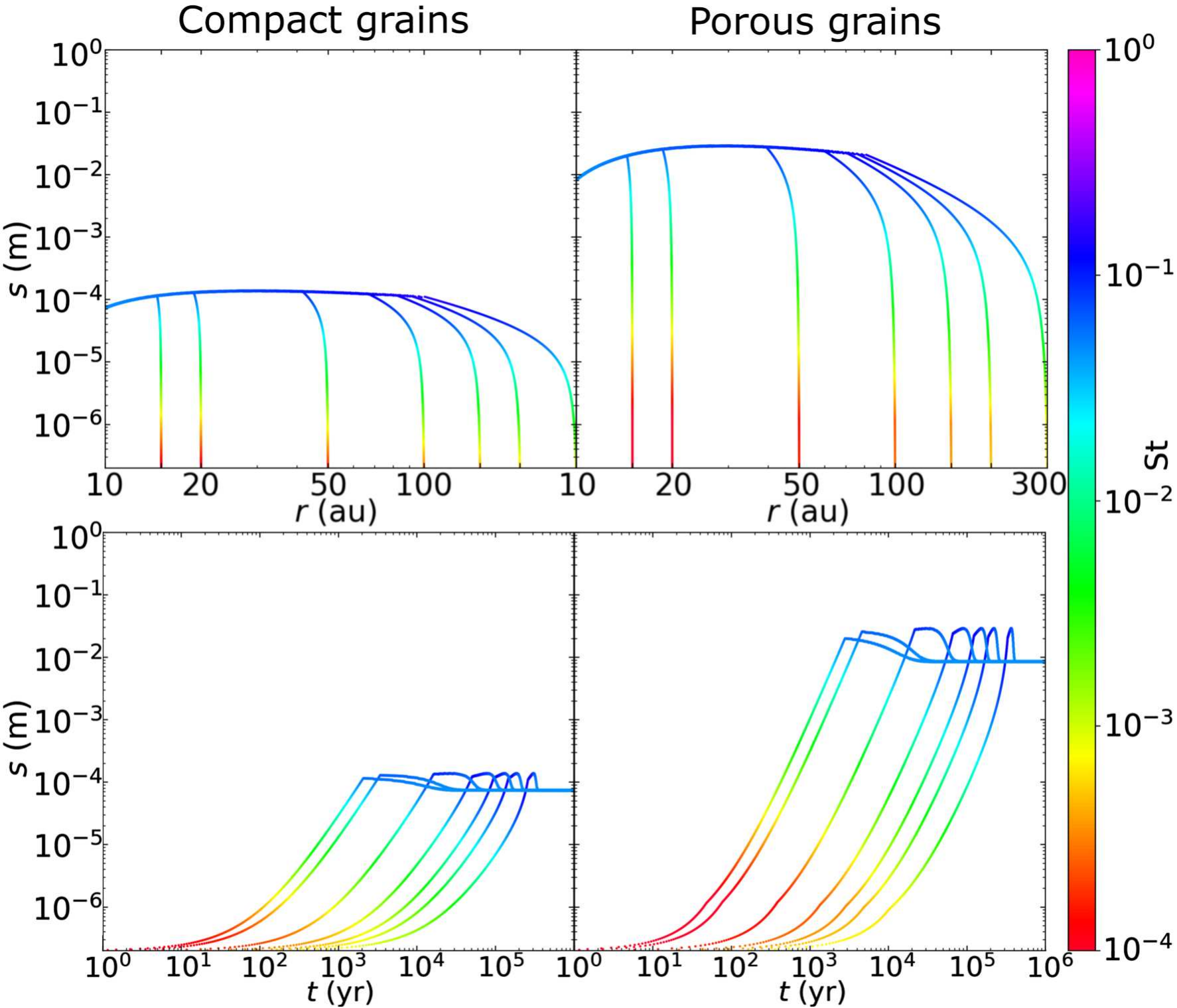}
  \caption{Comparison between \textsc{Pamdeas} simulations GF-Si-comp-Vf20  (left) and GF-Si-a02-Vf20 (right). The top panels show the size $s$
  as a function of distance $r$, with colour indicating the Stokes number St. The bottom panels show the size $s$
  as a function of time $t$, with colour indicating the Stokes number St as well. The leftmost curve corresponds to the grain initially at 15~au, and the rightmost to 300~au.}
  \label{Fig:1D-Comp-GF-Si-comp/a02-Vf20}
\end{figure}

First, we will examine the effect of porosity using the 1D code \textsc{Pamdeas}. Here the disc is static, and only the evolution of grains is taken into account.
Figure~\ref{Fig:1D-Comp-GF-Si-comp/a02-Vf20} makes it easy to observe the influence of porosity between the GF-Si-comp-Vf20 
simulations on the left and GF-Si-a02-Vf20 on the right.
The upper left (right) panel shows the size of compact (porous) silicate grains as a function of the distance to the star $r$. 
The colour indicates the coupling with the gas through the Stokes number. Each grain grows from the monomer size without drifting 
because it is coupled to the gas (vertical lines). The grains start drifting inwards when $\mathrm{St}\sim5\times 10^{-3}$, and they continue to grow 
until they reach the fragmentation threshold (horizontal plateau), maintaining an equilibrium between growth and fragmentation.
While the Stokes numbers (computed as per Eq.~5 in \citealt{vericel_dust_2021}) are similar in both simulations, with $\mathrm{St}\sim 10^{-1}$ when the grains reach the fragmentation threshold, the sizes are 
drastically different. Compact grains reach about 100 $\mu$m, whereas porous grains reach about 3 cm. These porous grains have a 
filling factor $\phi \sim 2\times 10^{-3}$, making them $\sim$50,000 times more massive than compact grains\footnote{To estimate the mass ratio between porous and compact grains, we can write the mass ratio as 
$m_\mathrm{porous}/m_\mathrm{compact}=\phi (s_\mathrm{porous}/s_\mathrm{compact})^3$.}.
Indeed, porous grains are capable of growing to larger sizes due to their larger cross-sectional area 
\citep{garcia_evolution_2020}. Moreover, porous grains remain well coupled to the gas even at sizes of 100 $\mu$m.

The bottom panels of Fig.~\ref{Fig:1D-Comp-GF-Si-comp/a02-Vf20} display the size evolution. Porous and compact grains reach their maximum size in similar timescales. 
A porous grain initially at 15~au grows to 3 cm in 1000 years, while a compact grain starting at the same location reaches 100~$\mu$m. Since the fragmentation 
threshold Stokes numbers are similar ($\mathrm{St}\sim0.1$), the aerodynamic evolution is the same, and hence the grains drift at the same speed. 
While drifting, the grains reach the inner regions where relative velocities are higher. This explains why the size plateaus have decreased to 8 mm for porous grains and 70 $\mu$m for compact grains at a distance of $r=10$ au, even though the grains were able to grow to larger sizes at a larger distance.

\begin{figure}
  \centering
  \includegraphics[width=\columnwidth,clip]{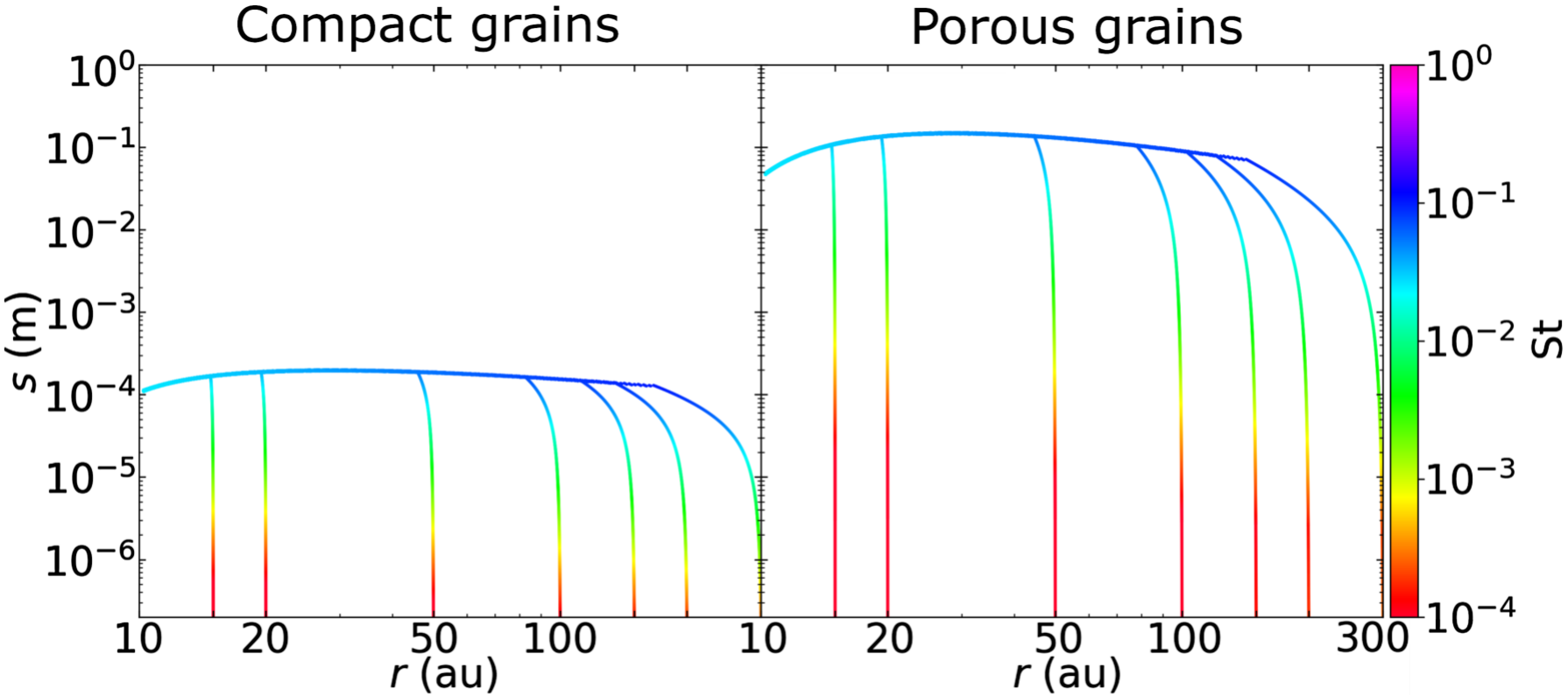}
  \caption{Same as the top panels of Fig.~\ref{Fig:1D-Comp-GF-Si-comp/a02-Vf20} for \textsc{Pamdeas} simulations GF-H2O-comp-Vf15 (left) and GF-H2O-a02-Vf15 (right).}
  \label{Fig:1D-Comp-GF-H2O-comp/a02-Vf15}
\end{figure}

A similar behaviour can be seen for the comparison between the GF-H2O-comp-Vf15 and GF-H2O-a02-Vf15 simulations on Fig.~\ref{Fig:1D-Comp-GF-H2O-comp/a02-Vf15}. 
Porous water ice grains grow to a size of 20 cm, while compact grains only grow to 200~$\mu$m. These sizes are larger than for silicate grains because of a lower intrinsic density, which increases the growth/fragmentation equilibrium size. At the fragmentation threshold, porous water ice grains are $5\times 10^{5}$ times more massive than compact grains.

In all cases, compaction by gas or self-gravity is never reached, 
as fragmentation maintains the grain sizes below a few metres, a range where gas compaction appears. Porous grains 
can therefore grow to much larger sizes (from centimetres to decimetres) and masses compared to their compact counterparts (a few hundreds of micrometres), demonstrating that porosity must be considered in dust evolution.

\subsubsection{3D global Simulations with \textsc{Phantom}}\label{Ssc:Porosity_Effects_3D_Simulations}

\begin{figure*}
  \centering
  \includegraphics[width=\textwidth,clip]{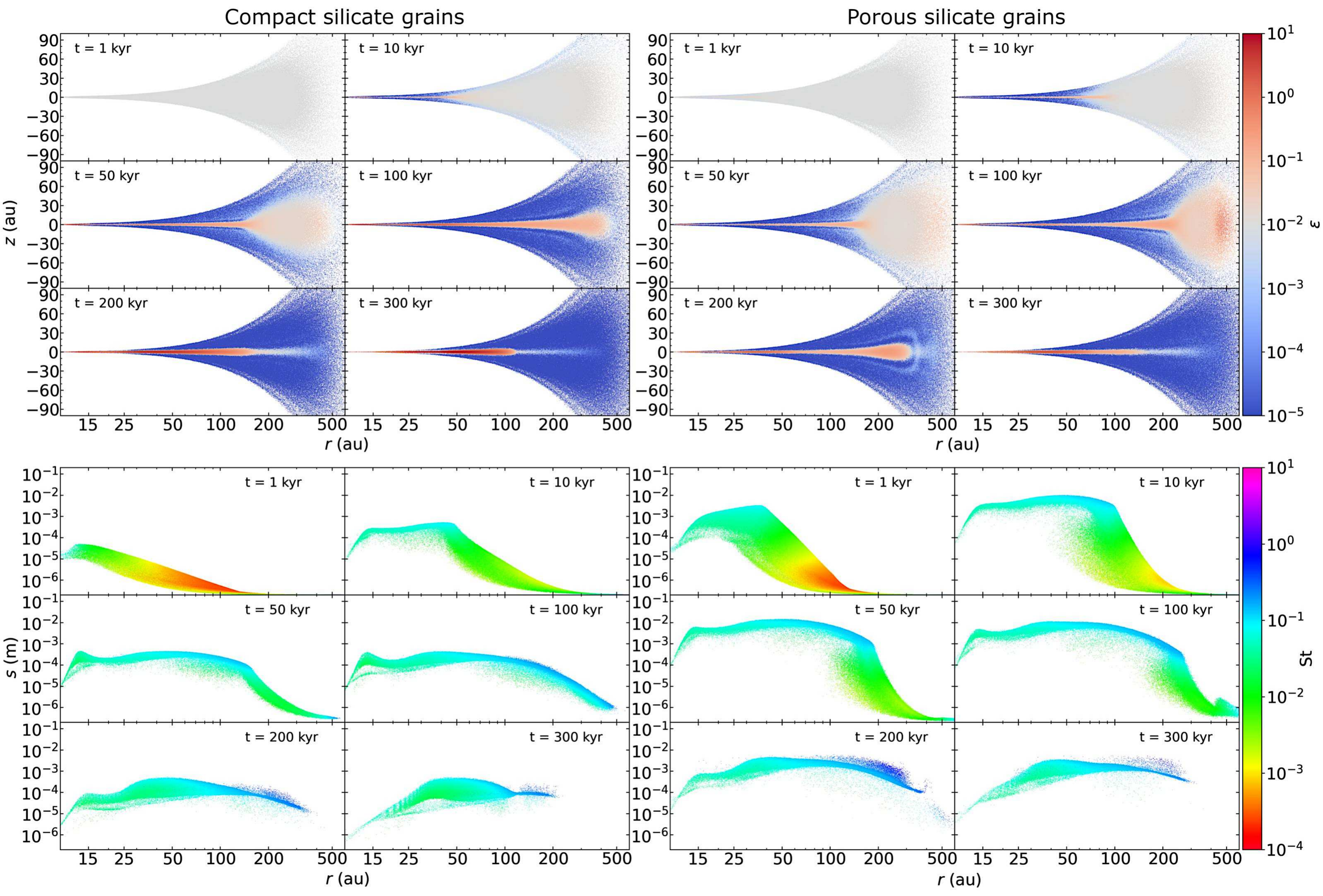}
   \caption{Comparison between \textsc{Phantom} simulations GF-Si-comp-Vf20 (left) and GF-Si-a02-Vf20 (right). 
   The top panels show the dust-to-gas ratio in colour in the ($r$,$z$) plane. 
   The bottom panels show the radial grain size distribution, with colour representing the Stokes number St.}
  \label{Fig:Comp-GF-Si-comp/a02-Vf20}
\end{figure*}

\begin{figure*}
 \centering
 \includegraphics[width=\textwidth,clip]{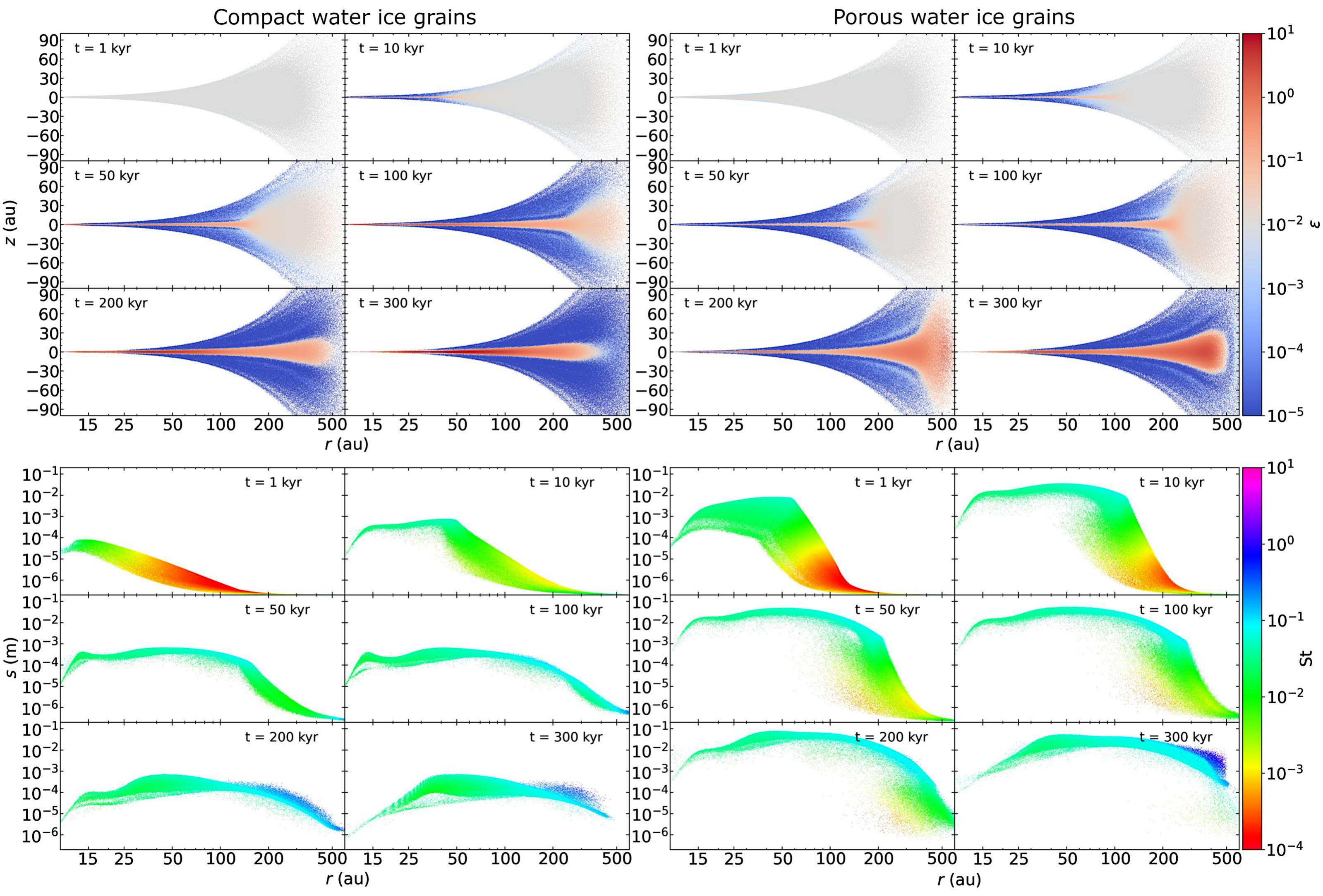}
  \caption{Same as Fig.~\ref{Fig:Comp-GF-Si-comp/a02-Vf20} for \textsc{Phantom} simulations GF-H2O-comp-Vf15 (left) and GF-H2O-a02-Vf15 (right), using water ice instead of silicate grains.}
  \label{Fig:Comp-GF-H2O-comp/a02-Vf20}
\end{figure*}

We now explore the effects of porosity with \textsc{Phantom}.
Figure~\ref{Fig:Comp-GF-Si-comp/a02-Vf20} compares the GF-Si-comp-Vf20 simulation (left) to GF-Si-a02-Vf20 (right). 
The top panels display the dust-to-gas ratio in the meridional plane ($r$, $z$) at different simulation times.
Gray areas represent the initial dust-to-gas ratio, red regions are dust-enriched, and blue regions are dust-depleted. 
It can be noticed that in the early times, porous grains settle faster in the midplane compared to compact grains, as they grow rapidly in the inner regions.
However, in the outer regions, porous grains, due to their low density, remain more coupled to the gas. 
They drift and settle more slowly, resulting in lower dust concentration and a thicker dust disc. This effect is evident at times $t=50$ or $100\ \mathrm{kyr}$.
At the end of the simulation, the compact dust disc is less radially extended ($\approx 100\ \mathrm{au}$) and 
has a higher dust concentration, around $\varepsilon \approx 3-4$, with a thickness of $\sim 10\ \mathrm{au}$.
In contrast, the porous dust disc extends out to 150 $\mathrm{au}$ with a dust-to-gas ratio of $\varepsilon \approx 0.8-1$. 
Porous grains thus allow retaining dust in the outer regions for a longer time while achieving large dust concentrations. Similarly to simulations with growth and fragmentation in \citet{garcia_evolution_2018}, the disc of porous grains after 300~kyr is thinner than that of compact grains.

The bottom panels of Fig.~\ref{Fig:Comp-GF-Si-comp/a02-Vf20} display the radial size distribution of dust grains. Colour represents the Stokes number St, 
where red to green grains are strongly coupled to the gas, blue grains are marginally coupled, and purple grains are decoupled.
To understand what happens in dust-concentrated regions, particles with $\varepsilon < 5\times 10^{-3}$ have been 
deliberately excluded, representing regions highly depleted in dust where little activity occurs.
Porous grains can grow to more substantial sizes, on the order of millimetres or centimetres, 
with filling factors $\phi\sim 5\times10^{-3}-10^{-2}$ (see Fig.~\ref{Fig:Phantom-Compaction-effect-r-s-phi}, left panel on second row), across a large portion of the disc, up to 200 $\mathrm{au}$.
In contrast, compact grains struggle to reach millimetre sizes and the largest ones remain in the 100-500 $\mathrm{\mu m}$ range.
This is inconsistent with observations that report the presence of millimetre sized grains in the midplane.
Hence, porosity helps grain growth to more significant sizes and masses, approximately 5 to 5000 times larger, 
with dust concentrations and disc thicknesses similar to compact grains, and sizes compatible with observations. 
However, the resulting filling factors (Fig.~\ref{Fig:Phantom-Compaction-effect-r-s-phi}) are too small by a factor of $\sim10-20$ compared to observations (see Sect.~\ref{Sc:Discussion}).
In both cases, grains do not cross $\mathrm{St} = 1$ and do not decouple from the gas.

Despite special attention to avoid numerical artefacts, some remain, inherent to the evolution model itself.
For porous grains at time $t=100\ \mathrm{kyr}$, a region close to 500~au where grains have grown slightly more than those closer in can be observed. This can be explained rather simply.
Initially, in the farthest regions (>300 $\mathrm{au}$), there are only monomers.
Monomers at 600 $\mathrm{au}$ have larger St values than those at 500 $\mathrm{au}$. They drift and settle 
faster than grains closer to the star can grow, resulting in a more dust-enriched zone, also visible in the top right panel.
All grains drift, settle, and grow, and this artefact related to the model disappears without consequence.

The same behaviour is seen for water ice grains, whose evolution is shown in Fig.~\ref{Fig:Comp-GF-H2O-comp/a02-Vf20}.
Compared to silicate grains, the top panels  demonstrate that water ice dust discs are thicker ($\sim 20\ \mathrm{au}$) and extend radially further (> 400 $\mathrm{au}$).  This is due to the lower density of water ice grains, resulting in smaller Stokes numbers for a given position and size. Water ice grains thus settle and drift more slowly than silicate grains. However, dust-to-gas 
ratios of around $\varepsilon \approx 2-3$ for compact grains and $\varepsilon \approx 0.7-1$ for porous grains are still reached.
Compact water ice grains struggle to reach millimetre sizes in the midplane (lower panels), although their growth is faster than that of compact 
silicate grains due to their lower density and larger cross-sectional area for a given mass. They remain relatively small, in the hundreds of microns range.
In contrast, porous water ice grains can easily attain centimetre scale sizes. Millimetre to centimetre sized grains are found out to 400-500 $\mathrm{au}$.
The resulting filling factors (see also Fig.~\ref{Fig:Phantom-Compaction-effect-r-s-phi}, bottom left), around $10^{-3}-5\times 10^{-3}$, are also too small compared to observations by a factor 20 to 100.
Grains are still maintained at St values around 0.1; none manage to decouple in the inner regions. 
This is not the case for some grains that reach $\mathrm{St} = 1$ in the outer disc region, where lower gas density and a higher growth rate enable reaching these St values.
Thus, porosity allows grains to grow more easily, reaching millimetre sizes, and forming dust discs as thin as those composed of compact grains.

\subsection{Effect of bouncing and compaction during fragmentation}\label{Ssc:compaction_effect}
In this section, we investigate the effects of compaction during fragmentation and bouncing at different fragmentation thresholds. 
Disc observations reveal that larger dust particles in the midplane are not as porous as the values obtained in the absence of compaction (Sect.~\ref{Ssc:porosity_effect}). 
\citet{Kataoka_polar_2016}, \citet{Guidi_2022}, and \citet{Zhang_2023} find that grains have filling factors on the order of 10\% with a product 
$s\phi$ larger than 100~$\mu$m, meaning grains with $s>1$~mm.

\begin{figure}
  \centering
  \includegraphics[width=\columnwidth,clip]{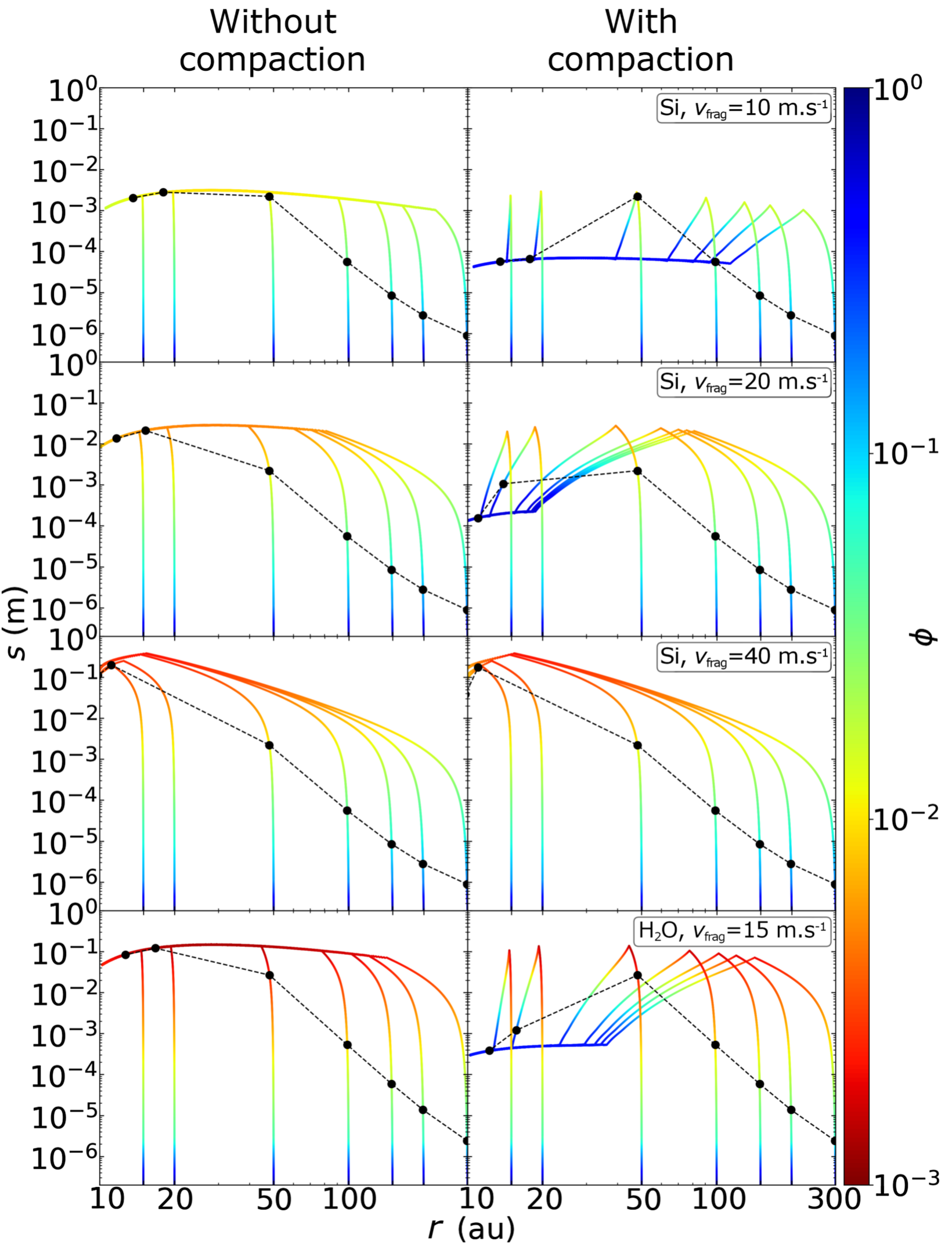}
  \caption{Comparison between \textsc{Pamdeas} simulations with growth and fragmentation (GF-*-a02-*, left) and simulations with growth, bouncing, and fragmentation with compaction (GBFc-*-a02-*, right), up to $t=1$~Myr. Grain sizes are given as a function of the distance from the star, with colour indicating the filling factor. Simulation parameters are indicated in each row. Black dots and dashed lines mark the time $t=50$ kyr, providing a `snapshot' of the size distribution at that moment.}
  \label{Fig:Pamdeas-Compaction-effect-r-s-phi}
\end{figure}

\subsubsection{1D Simulations with \textsc{Pamdeas}}
\label{Ssc:Compaction_Effect_1D_Simulations}

Figure~\ref{Fig:Pamdeas-Compaction-effect-r-s-phi} compares grain sizes as a function of distance from the star 
between simulations without bouncing and compaction during fragmentation (GF) on the left, and those with (GBFc) on the right. The top three rows correspond to simulations with silicates for three fragmentation thresholds: $v_\mathrm{frag,\:Si} = 10$, 20, and 40 $\msec$, from top to bottom, while the bottom row shows simulations with water ice for $v_\mathrm{frag,\:H2O} = 15\ \msec$. For all GBFc simulations, the grains 
grow to the same sizes and filling factors as the GF simulations. However, when the fragmentation threshold is reached, 
the grains in the left panels remain in an equilibrium between growth and fragmentation because, during fragmentation, the grains 
retain their filling factor. In the right panels, aggregates are compacted during fragmentation, the filling factor increases, 
and sizes become smaller until a new equilibrium is reached. The value of $\phi$ for compacted grains is not the maximum possible 
value $\phi_\mathrm{max}= 0.74$. The maximum value that is reached is instead around $0.5-0.6$ as in equilibrium, grain growth tends to decrease $\phi$, 
while compaction by fragmentation or bouncing tends to increase $\phi$. According to Fig.~\ref{Fig:fig_croissance_fragcomp_rebond_pamdeas_drift} 
and Eq.~(\ref{Eq:Pcomp}), the closer $\phi$ gets to the maximum value, the harder it is to compact the aggregate. 
The slow growth of grains thus compensates for compaction towards the maximum value.

For simulations with $v_\mathrm{frag,\:Si} = 10\ \msec$, 
the grains reach sizes of a few millimetres, with a maximum around 3 mm between 20 and 50~au. When compacted, the aggregate sizes then 
drop to about $40-80\ \mu$m at maximum compaction before being accreted by the star. The filling factor reaches a minimum of $10^{-2}$ 
just before the fragmentation threshold, before being compacted to $\phi = 0.5$. The aggregates are compacted very efficiently, as a grain 
initially growing at 300 au ends up fully compacted during its drift beyond 100 au. Lastly, one can note that the compacted grains 
drift as fast as those that do not undergo compaction, as indicated by the black points and line. This is due to the fact that the product $s\phi$ 
during compaction remains roughly the same as when there is no compaction. As $\mathrm{St} \propto s\phi$, a similar Stokes number 
results in a similar drift.

With $v_\mathrm{frag,\:Si} = 20\ \msec$, grains in the right panel reach sizes 
of around 3~cm before being compacted. As the fragmentation threshold is higher, the grains have larger Stokes numbers (St) and 
therefore drift more during compaction. The maximum compaction occurs between 10 and 20 au, resulting in sizes ranging from 100 to 
300 $\mu$m and filling factors between 0.3 and 0.5.

For $v_\mathrm{frag,\:Si} = 40\ \mathrm{\msec}$, grains are able to grow almost without encountering 
the fragmentation barrier. In both cases, the grains start to fragment when reaching sizes between 20 and 50 cm, with filling factors 
smaller than $3 \times 10^{-3}$. With such a high fragmentation threshold, grains undergo mostly pure growth in the majority of the disc and compaction does not operate. Moreover, due to this free growth, grains in the inner regions are accreted more rapidly since the $\mathrm{St}$ values they reach are larger, around $0.5-1$ between 10 and 20 au, and over 0.1 in the rest of the disc.

Finally, in the case of water ice (bottom row of Fig.~\ref{Fig:Pamdeas-Compaction-effect-r-s-phi}), 
the aggregates reach decimetre sizes in both cases. When compacted, their size is reduced by a 
factor of 500. The size reached at maximum compaction ranges from 300 to 500~$\mu$m, with a filling factor $\phi \sim 0.4$. Thus, a behaviour similar to simulations with silicates and $v_\mathrm{frag,\:Si} = 20\ \mathrm{\msec}$ is 
observed, but with larger variations in size and filling factor.

\begin{figure}
   \centering
   \includegraphics[width=\columnwidth,clip]{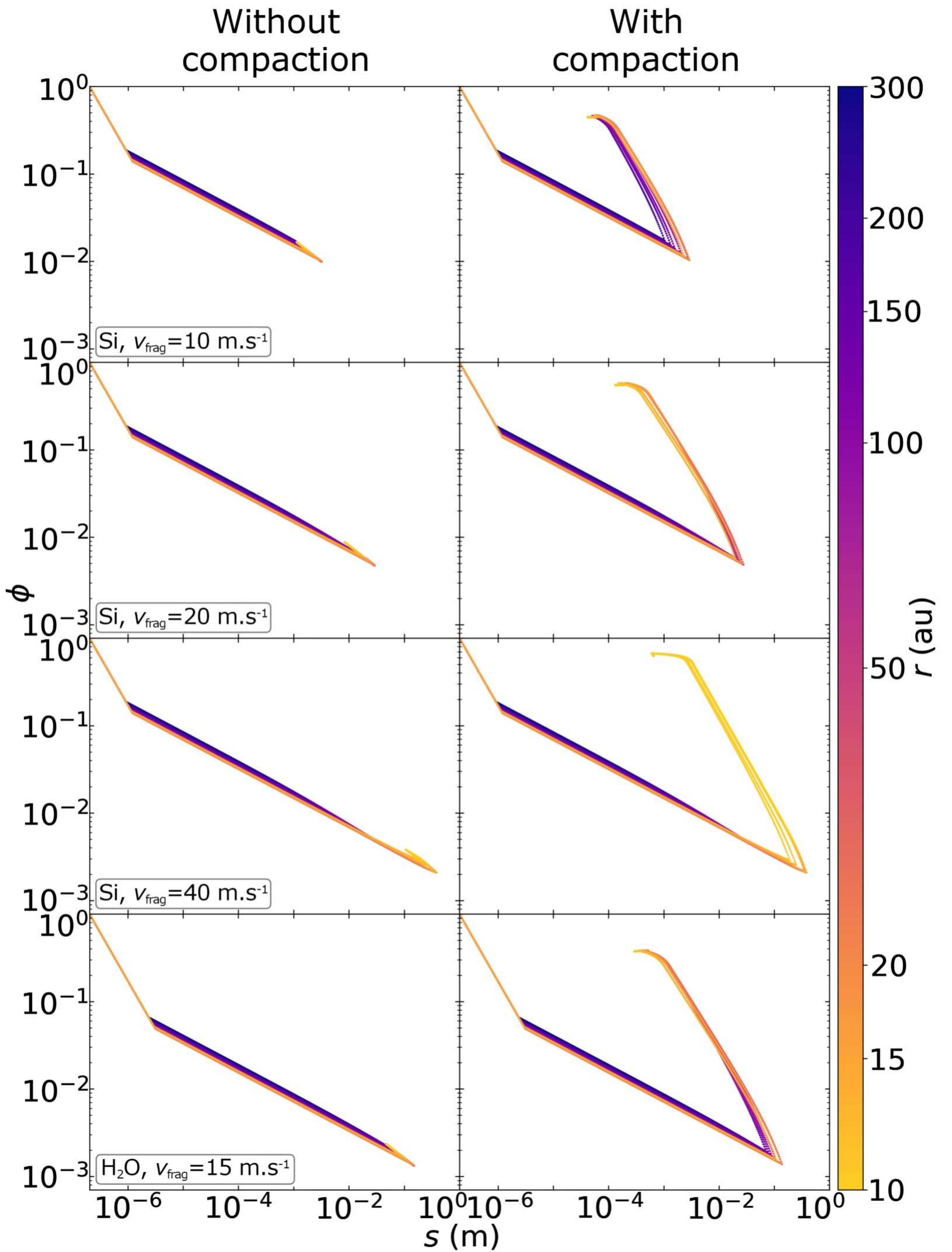}
   \caption{Comparison between \textsc{Pamdeas} simulations with growth and fragmentation (GF-*-a02-*, left) and simulations with growth, bouncing, and fragmentation with compaction (GBFc-*-a02-*, right), up to $t=1$~Myr. The filling factor $\phi$ is plotted against grain size, with colour indicating the distance $r$. Simulation parameters are indicated in each row.}
   \label{Fig:Pamdeas-Compaction-effect-s-phi-r}
\end{figure}

Figure~\ref{Fig:Pamdeas-Compaction-effect-s-phi-r} shows the evolution of size and filling factor, with colour representing the distance $r$. All grains start to grow in the hit \& stick regime from the size of a monomer $a_0 = 0.2\ \mu$m and $\phi = 1$. They then transition to a different growth regime ($\phi_\mathrm{Ep,\ St<1}$) resulting in a change in slope around the micrometre size, depending on the species (see Fig.~\ref{Fig:fig_croissance_fragcomp_rebond_pamdeas_drift}).

In left panels for simulations without compaction, once the fragmentation threshold is reached, the grains remain confined in the `tip' at the bottom right. On the other hand, in the right panels, grain compaction is observed, leading to a decrease in size while the filling factor increases towards unity. With $v_\mathrm{frag,\:Si} = 10\ \mathrm{\msec}$, grains are compacted up to $\phi = 0.5$ and a minimum size of 40~$\mu$m, even at larger distances $r$. Bouncing doesn't occur in this case, as fragmentation is highly effective. For $v_\mathrm{frag,\:Si} = 20\ \mathrm{\msec}$, grains fragment at larger sizes and smaller $\phi$. However, once the fragmentation threshold is reached, they are compacted up to $\phi = 0.6$ and sizes of 100 $\mu$m. Bouncing begins to appear but has no influence on grain evolution and is not visible in the figure. For $v_\mathrm{frag,\:Si} = 40\ \mathrm{\msec}$, aggregates fragment after growing to sizes of around a decimetre and $\phi\sim2\times10^{-3}$. They then fragment and are compacted to $\phi = 0.7$ and sizes of 700 $\mu$m.
   
Finally, in the case of water ice, the same behaviour is observed. Aggregates are compacted until they reach $\phi = 0.4$ and $s = 300\ \mu$m. As mentioned in Sect.~\ref{Ssc:bounce}, bouncing is accompanied by growth (but hardly visible in Fig.~\ref{Fig:Pamdeas-Compaction-effect-s-phi-r}). Similarly to Fig.~\ref{Fig:fig_croissance_fragcomp_rebond_pamdeas_drift}, the characteristic plateau is seen, where grains move from left to right and bottom to top, as they are compacted by bouncing during growth before continuing to drift and be accreted.
   
Thus, compaction during fragmentation has a significant impact on grain evolution. Taking porosity into account allows grains to overcome the bouncing barrier, which occurs at grain sizes of a few to several tens of microns. 
Bouncing doesn't interfere with aggregate growth before reaching the fragmentation threshold, as $\phi < 0.3$. When $\phi$ becomes larger than 0.3 again, grains grow and don't bounce but fragment. 
It is necessary for the grains to be compacted and at sufficiently small sizes for bouncing to play a role again before fragmentation.

\begin{figure*}
  \centering
  \includegraphics[width=\textwidth,clip]{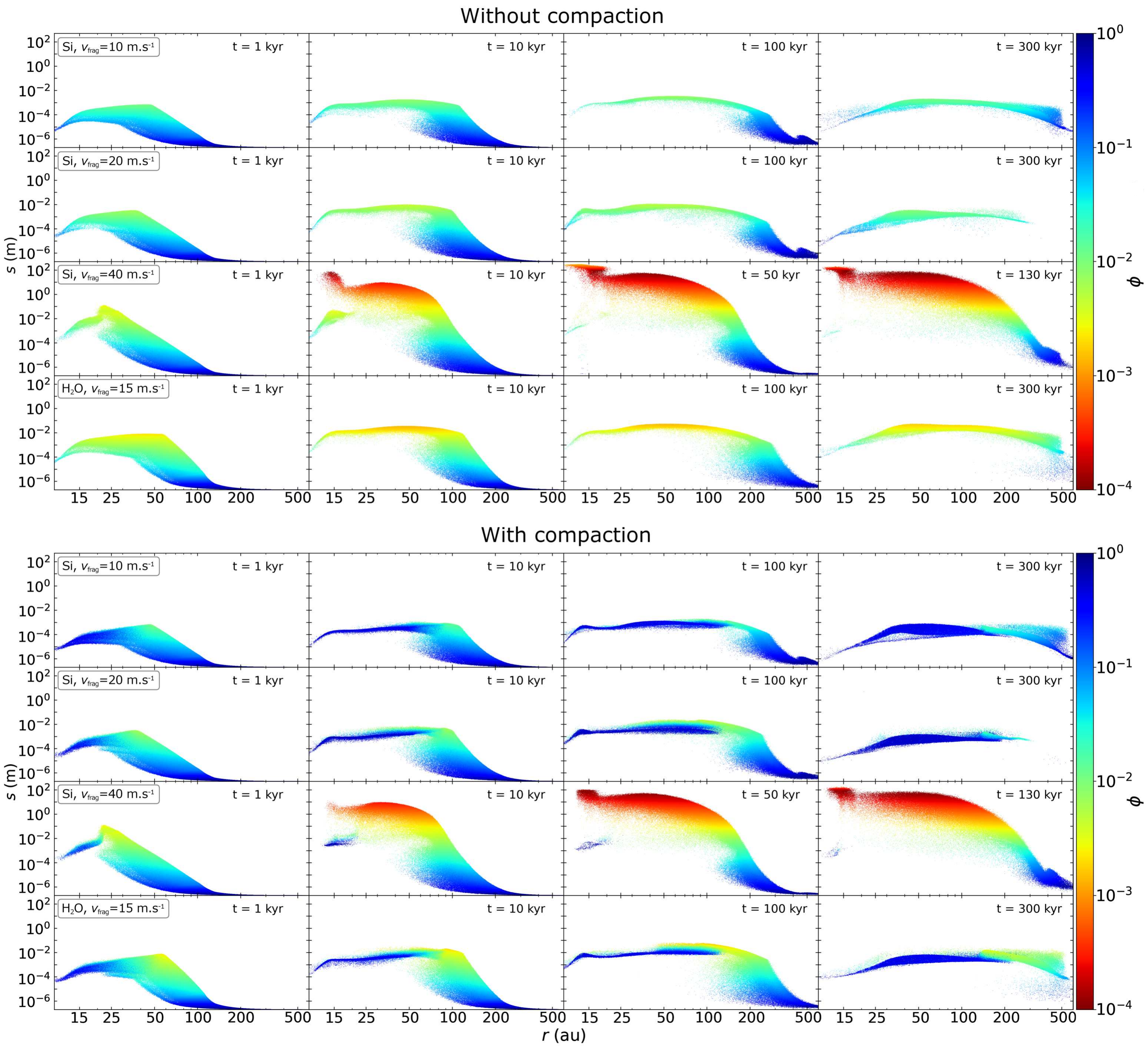}
  \caption{Comparison of the radial grain size distribution, with colour representing the filling factor, between \textsc{Phantom} simulations with growth and fragmentation (GF-*-a02-*, top panel) and with growth, bouncing, and fragmentation with compaction (GBFc-*-a02-*, bottom panel). Simulation parameters are indicated in each row. Note that the colorbar range is different from that in Fig.~\ref{Fig:Pamdeas-Compaction-effect-r-s-phi}.}
  \label{Fig:Phantom-Compaction-effect-r-s-phi}
\end{figure*}

\begin{figure*}
   \centering
    \includegraphics[width=\textwidth,clip]{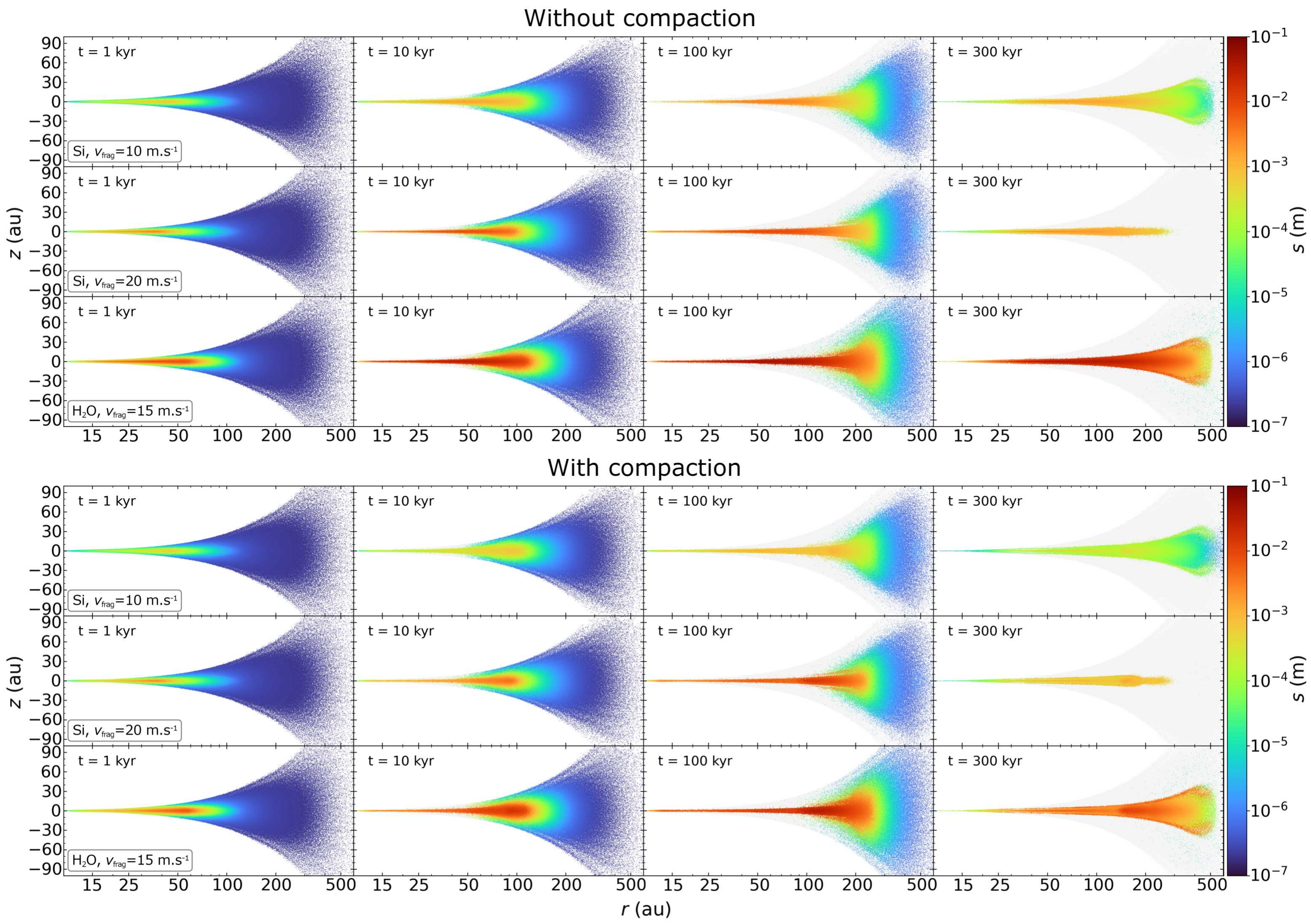}
   \caption{Comparison between \textsc{Phantom} simulations with growth and fragmentation (GF-*-a02-*, top panel) and simulations with growth, bouncing, and fragmentation with compaction (GBFc-*-a02-*, bottom panel).
   The grain size $s$ is shown in the meridian plane ($r$, $z$), with the light gray background indicating the gas disc thickness.
   Simulation parameters are indicated in each row.}
   \label{Fig:Phantom-Compaction-effect-r-z-s}
\end{figure*}

\subsubsection{3D global Simulations with \textsc{Phantom}}\label{Ssc:Compaction_Effect_3D_Simulations}

Figure~\ref{Fig:Phantom-Compaction-effect-r-s-phi} compares \textsc{Phantom} simulations without (top) and with (bottom) compaction during fragmentation and bouncing. 
We show the radial grain size distribution, colour-coded with the filling factor, for simulations with silicates and $v_\mathrm{frag,\:Si} = 10$, 20 and $40\ \msec$, and with water ice, from top to bottom in each panel (in the same order as in Fig.~\ref{Fig:Pamdeas-Compaction-effect-r-s-phi}). In all cases, the maximum grain sizes obtained with or without compaction are similar, unlike in the 1D simulations, despite efficient compaction. It should be noted that 1D and 3D simulations cannot be directly compared as the former follow the evolution of single grains in the disc midplane while the latter deal with a population of grains at various altitudes, whose size and porosity evolution depends on varying disc conditions as they settle and drift. The effect of compaction (bottom panel) can best be seen at $t=100$~kyr interior to 200~au where the largest grains are still relatively porous, with $\phi$ ranging from a few $10^{-2}$ to a few $10^{-3}$ depending on the simulation, and mostly at large $r$, while the smallest grains have been compacted to $\phi\sim0.4$ or larger. In simulations without compaction (top panel), the larger grains are also more porous than smaller ones, as expected from their evolution during growth (Sect.~\ref{Sc:Methods}),  but the porosity range is much smaller. At $t=300$~Myr, the most compacted grains, in the midplane, are substantially smaller than in simulations without compaction.

Simulations with silicate grains and $v_\mathrm{frag,\:Si} = 10$ and 20~$\msec$ (first two rows in both panels) are similar: without or with compaction, grains reach sizes of a few mm at $t=100$~kyr for $v_\mathrm{frag} = 10$~$\msec$, while they grow faster and larger, up to a few cm, for $v_\mathrm{frag,\:Si} = 20$~$\msec$. However, after 300~kyr, compacted grains are at most 600-700 $\mu$m in size for $v_\mathrm{frag} = 10$~$\msec$ and a few mm for $v_\mathrm{frag,\:Si} = 20$~$\msec$. The former case is difficult to reconcile with observations of grains of mm size or larger in protoplanetary discs.
When $v_\mathrm{frag,\:Si} = 40$~$\msec$ (third rows), the threshold is high enough for grains to remain mostly in the pure growth regime, similarly to what was seen with \textsc{Pamdeas} (Sect.~\ref{Ssc:Compaction_Effect_1D_Simulations}), with almost no fragmentation or compaction, except in the very inner disc. Here, grain sizes reach several tens of metres and their Stokes number exceeds unity. This breaks the terminal velocity approximation used in the dust-as-mixture formalism \citep{youdin_streaming_2005,Laibe_2014a} and those simulations should not be considered valid.
Finally, for simulations GF-H2O-a02-Vf15 and GBFc-H2O-a02-Vf15 (bottom rows), the same pattern as for simulations GF-Si-a02-Vf20 and GBFc-Si-a02-Vf20 is observed, with grains reaching slightly larger sizes. The joint evolution of grain size and filling factor is described in Appendix~\ref{App:Phantom-Compaction-effect-s-phi-r}.

The thickness of the of the dust discs can be examined in Fig.~\ref{Fig:Phantom-Compaction-effect-r-z-s}, showing the size in the ($r$, $z$) plane. Like in other figures, only particles with $\varepsilon \geq 5 \times 10^{-3}$ are shown to eliminate dust-depleted regions. The light gray background indicates the gas disc's thickness. An interesting result can be noted: whether the grains are compacted or not, the thickness of the dust discs is very similar between 10 and 200 au, where dust is most abundant, and is only slightly larger for compacted dust. In all cases, compared to the compact grain discs of simulations GF-Si-comp-Vf20 (Fig.~\ref{Fig:Comp-GF-Si-comp/a02-Vf20}) and 
GF-H2O-comp-Vf15 (Fig.~\ref{Fig:Comp-GF-H2O-comp/a02-Vf20}), the porous grain discs are just as thin in the inner regions, or even thinner in the early stages for $r<200-300$ au. However, the porous grain discs are thicker farther from the star and have larger sizes than their compact grains counterparts.

Finally, the maximum size of compact grains is larger in \textsc{Phantom} simulations compared with \textsc{Pamdeas} by a factor of $\sim 5$ (comparing the left panels of Figure~\ref{Fig:1D-Comp-GF-Si-comp/a02-Vf20} to the left panels of \ref{Fig:Comp-GF-Si-comp/a02-Vf20}) because dust settling in 3D increases dust density in the midplane, which helps grains to grow to larger sizes. On the other hand, the sizes, filling factors, and St values of porous grains are similar with both codes. When grains are highly porous, their growth is fast and the increase in growth rate due to settling provides little assistance --- the limiting factor is the fragmentation threshold. Porous grains that undergo compaction are in a intermediate situation.

\subsection{Effects of a snow line}
\label{Ssc:snow_line_effect}

\begin{figure}
   \centering
   \includegraphics[width=\columnwidth,clip]{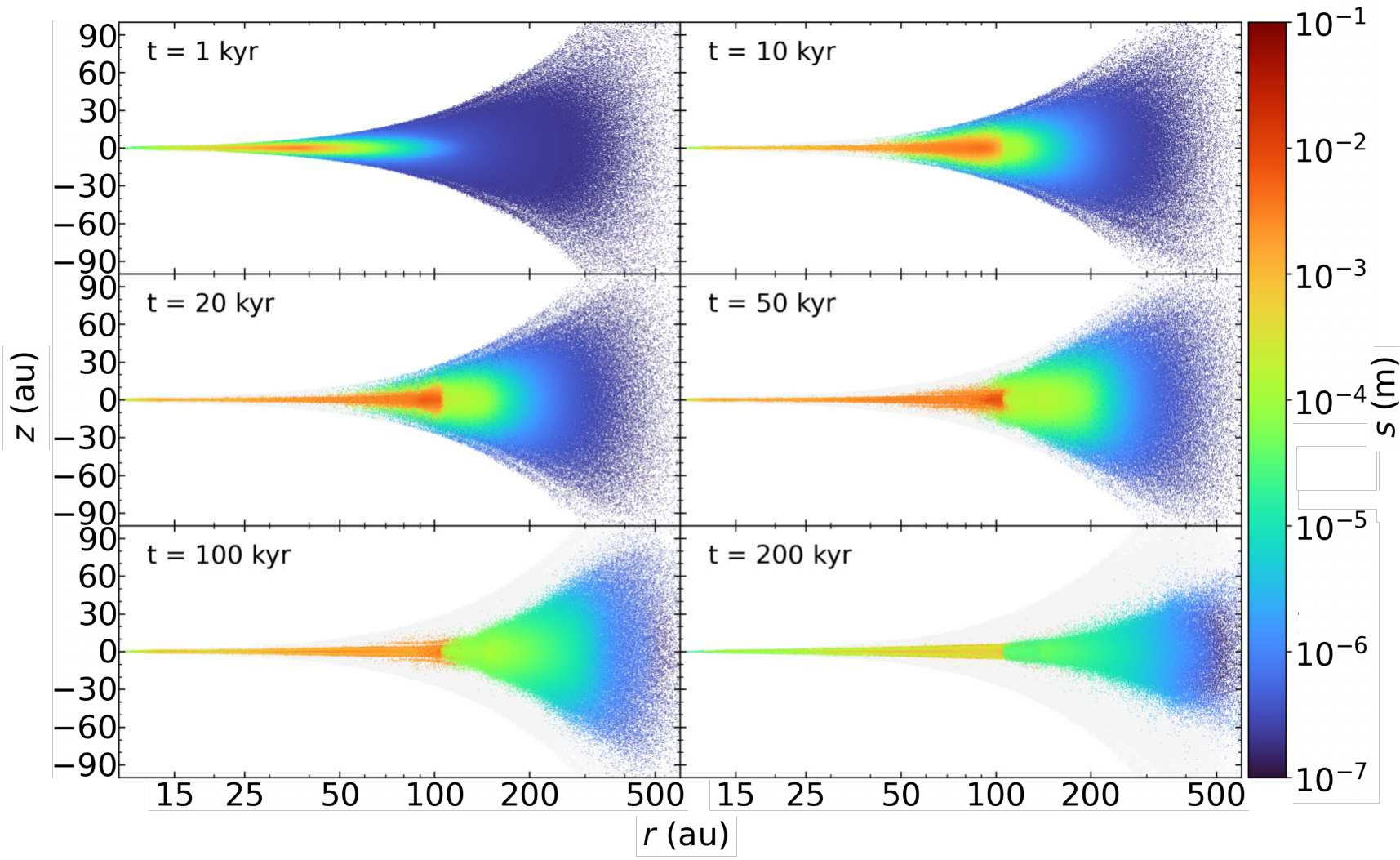}
   \includegraphics[width=\columnwidth,clip]{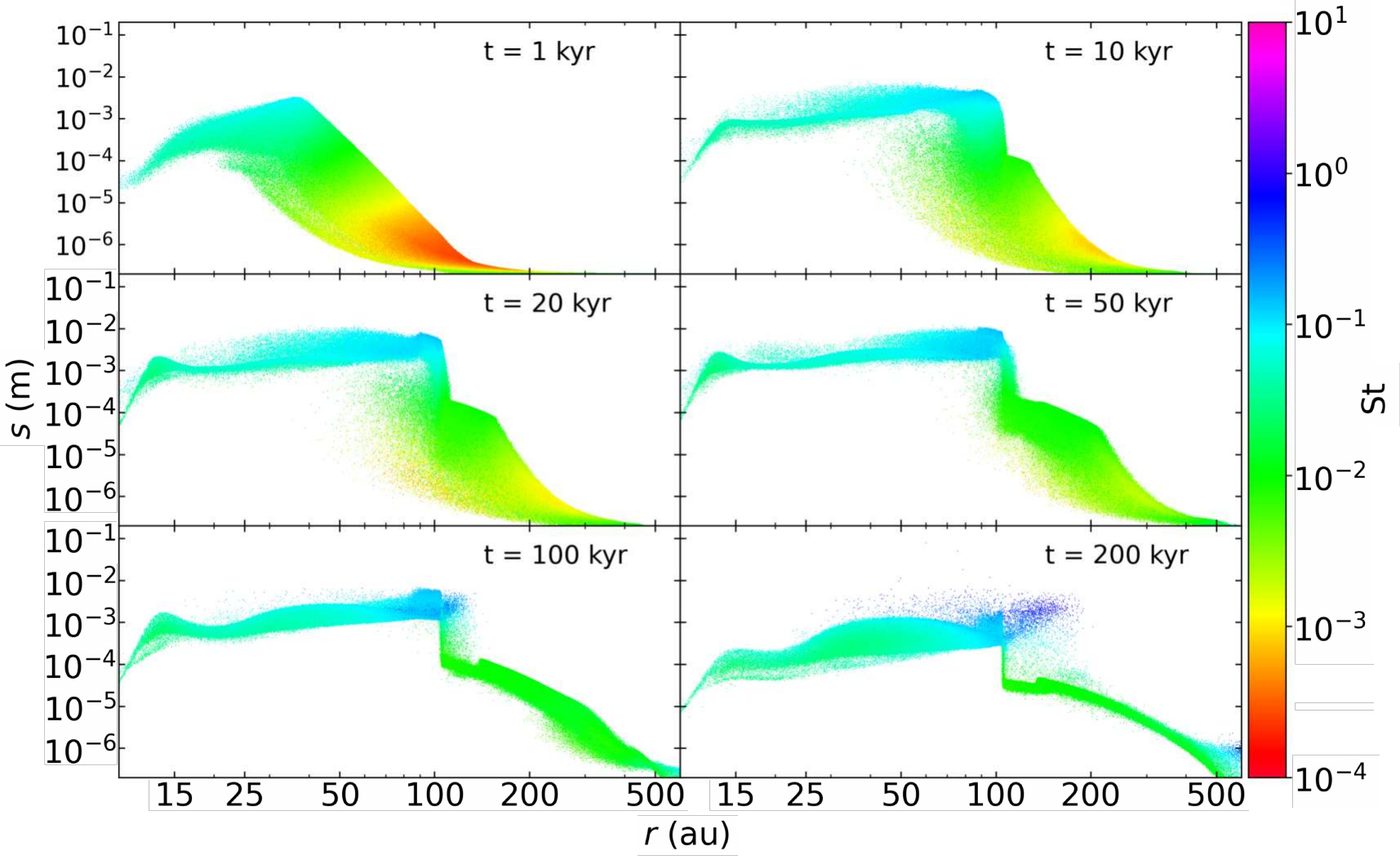}
   \includegraphics[width=\columnwidth,clip]{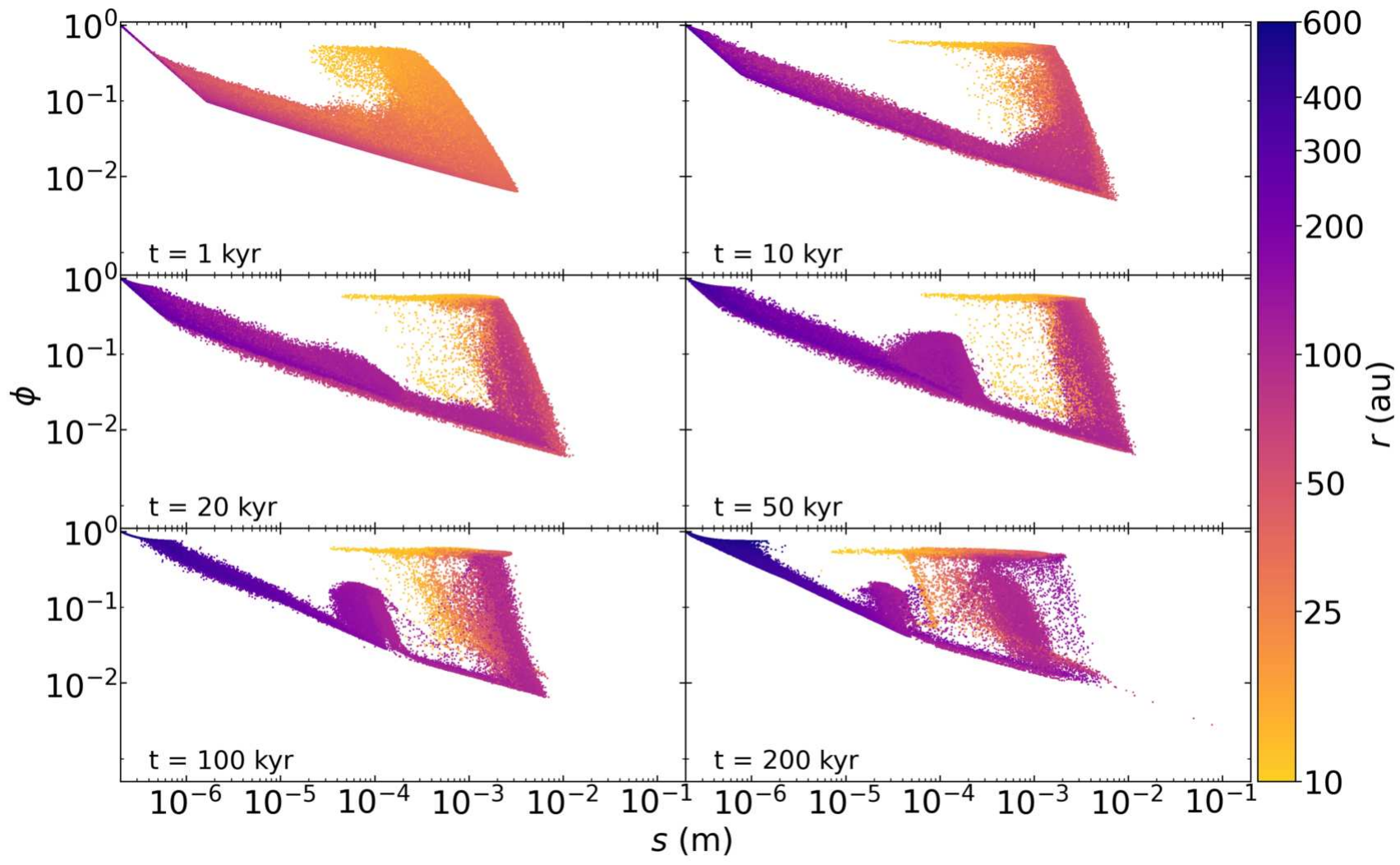}
   \caption{\textsc{Phantom} GBFcS-Si-a02-Vf205 simulation. \textbf{Top}: aggregate sizes in the same plane ($r$, $z$).
   \textbf{Middle}: radial grain size distribution, with colour representing the Stokes number St. \textbf{Bottom}: filling factor $\phi$ plotted against grain size $s$, with colour indicating the distance $r$.
   }
   \label{Fig:Phantom-1F-GBFcS-Si-a02-Vf205}
\end{figure}

Here, we examine the influence of the CO ice line on the GBFc-Si-a02-Vf20 simulation. The sublimation temperature for CO is 20 K, which corresponds to a position of the snow line at $r\sim 100$ au in our disc model. We set the fragmentation threshold for the outer region, where grains are assumed to be surrounded by CO ice, to $v_\mathrm{frag,\:ext}=5\ \msec$. In the inner region, instead of taking a threshold of 15~$\msec$, which would correspond to grains made entirely of water ice, we set $v_\mathrm{frag,\:int}=20\ \msec$ to model silicate grains assumed to be surrounded by a layer of water ice. The outer layer only affects adhesive properties, as discussed in \citet{vericel_self-induced_2020}. The result from this simulation, GBFc-Si-a02-Vf205, are shown in Fig.~\ref{Fig:Phantom-1F-GBFcS-Si-a02-Vf205}.

The top panel shows the grain size $s$ in the ($r$, $z$) plane, with a cut-off at $\varepsilon \geq 5 \times 10^{-3}$, and the light gray gas disc in the background, while the central panel displays the radial grain size distribution coloured by the Stokes number. Both panels show two distinct regions. Interior to 100~au, grains have settled into a very thin disc and formed large aggregates with sizes of several millimetres up to a centimetre. They have $\mathrm{St}\sim 5\times 10^{-2}$ when they are compacted, mainly between 50 and 100 au, down to a few mm. Beyond 100~au, grains have not been able to grow to large sizes due to the lower fragmentation threshold exterior to the snow line, where the balance between growth and fragmentation keeps them at smaller sizes ranging from $\sim100$~$\mu$m when they are still porous, down to a few tens of $\mu$m once compacted. Their smaller St slows down their settling and drift. The dust disc has thus settled much less compared to simulation GBFc-Si-a02-Vf20 shown in Fig.\ref{Fig:Phantom-Compaction-effect-r-z-s}, with a thicker and more extended outer disc. 

Finally, the bottom panel shows the filling factor versus the grain size $s$, with colour indicating the distance $r$. With the snow line, the interpretation becomes more complex, as grains can fragment and compact in two different regions, leading to compaction even for small sizes in the outer regions. Grains between 100 and 150~au are compacted, and once past the snow line, they have the opportunity to grow again before reaching the inner fragmentation threshold of 20~$\msec$. Thus, we observe not one but two vertical columns in the plot due to compaction during fragmentation.

Snow lines in this configuration are an effective way to form relatively compact dust grains with $\phi\sim 0.1$ far from the star, as long as the fragmentation threshold is not too high; otherwise, the grains would drift inward faster and would not be compacted efficiently.

\subsection{Effects of rotational disruption}
\label{Ssc:effect_rotational_disruption}

We present in Appendix~\ref{App:influence_of_rotational_disruption} the results of 3D simulations of rotational disruption, complementing the 1D \textsc{Pamdeas} simulations of \citet{Michoulier_Gonzalez_Disruption} and removing some of their limitations. They show that rotational disruption has a negligible effect on the dust evolution in discs.

\section{Discussion}\label{Sc:Discussion}

We have shown that simulations with porous grains allow the growth of grains to larger sizes and masses, regardless of the species or fragmentation threshold. Hence, as indicated by \citet{garcia_evolution_2020}, porosity enables large dust grain formation. Our 1D simulations using \textsc{Pamdeas} and our 3D simulations with \textsc{Phantom} both result in cm-sized grains for porous grains when including growth and fragmentation only. With compact or compacted grains due to fragmentation, \textsc{Pamdeas} tends to underestimate the maximum grain size by a factor of a few. Nevertheless, \textsc{Pamdeas} efficiently models the dust evolution, with or without considering porosity evolution.

We found that simulations involving compaction contradict the earlier simulations without compaction, which indicated that larger grains were more porous and located in the midplane \citep{garcia_evolution_2018, garcia_evolution_2020}. With compaction, we find that large aggregates in the midplane are compact. In the upper layers, grains are nearly as large but more porous and consequently less massive, undergoing settling.
Finally, small grains are either monomers or fractal aggregates composed of a few dozen monomers. This suggests that the size and porosity distribution of dust mainly depends on grain altitude.
Our simulations with porous and compacting grains provide better explanations for observations that report both non-porous millimetre sized grains in the midplane and more porous millimetre sized grains above and below.

\begin{figure}
   \centering
   \includegraphics[width=\columnwidth,clip]{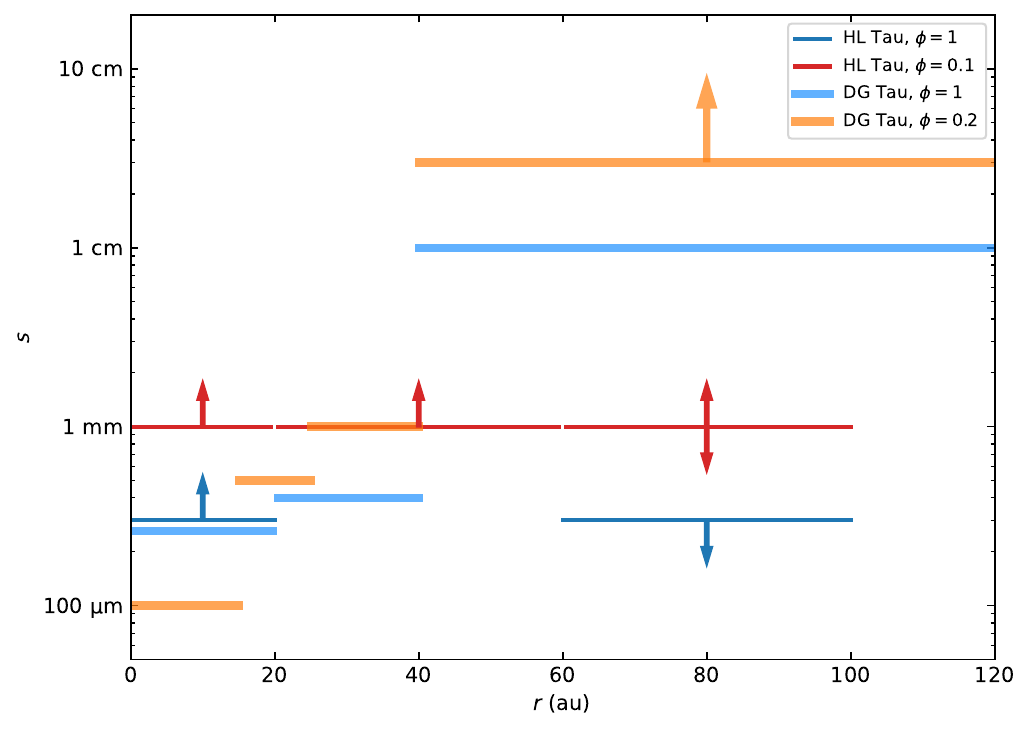}
   \caption{Grains sizes compatible with all the data obtained from SED and polarization observations of HL Tau \citep{Zhang_2023} and DG Tau \citep{Ohashi_2023} for different values of the assumed filling factor.}
   \label{Fig:Obs_constraints}
\end{figure}

Filling factors of dust grains in discs are generally between 0.1 and 1 according to observations \citep{Guidi_2022,Zhang_2023,Ohashi_2023}. Figure~\ref{Fig:Obs_constraints} summarizes the best models able to match observations of HL Tau \citep{Zhang_2023} and DG Tau \citep{Ohashi_2023}. In both studies, the authors assume either compact ($\phi=1$) of porous ($\phi=0.1$ for \citealt{Zhang_2023} and 0.2 for \citealt{Ohashi_2023}) grains and fit for the grain sizes for which SED and polarization data are best reproduced.
In the inner disc (interior to 40~au), compact grains of a few hundred $\mu$m are found to match both discs. However for porous grains, sizes of 1~mm or larger are preferred for HL Tau, while values from 100~$\mu$m closer in up to 1~mm further out work best for DG Tau. For HL Tau, \citet{Zhang_2023} found that compact grains of any size cannot reproduce the data between 20 and 60~au. At larger distances, the best fit is found for similar sizes to the inner disc for HL Tau, but for compact grains of 1~cm or porous grains larger than 3~cm for DG Tau. However, the authors of both studies caution that the low signal-to-noise ratio in the outer regions makes it difficult to distinguish between different grain populations.

Our simulations provide an explanation for these grain distributions. Thanks to porosity, all grains are capable of rapid growth to reach millimetre sizes that are compacted in the inner regions with consistent filling factors ($\phi\sim 0.1-1$). In the outer regions, smaller grains ($10-100\ \mu$m) are prevalent, with varying porosities ($\phi\sim 0.01-0.5$), which is consistent with the observations of HL Tau. The large grains ($\gtrsim 3$~cm) seen in the DG Tau disc are however challenging to reproduce in our simulations. However, grains with these sizes are present, but they are more porous with $\phi\sim 10^{-3}-10^{-2}$.

Table~\ref{Tab:guttler_2019} presents a summary of the classifications of cometary dust and its morphology from the Rosetta and Stardust missions \citep{guttler_synthesis_2019}. These observations found that the most common grains are represented by the porous particle group, with $\phi$ in the [0.05,0.9] range, similar to the compacted grains in our simulations, although their size can be much larger. Fractal aggregates are rarer, also in agreement with our simulations. The solid particle group is similar to our monomers. Compaction of grains during their evolution is thus necessary to explain both disc and comet observations.


Some limitations of our work are inherent to the codes themselves (see Sect.~\ref{Sc:Methods}), in particular for \textsc{Pamdeas} with its one-dimensional nature assuming a static gas disc, limitations that are overcome with \textsc{Phantom}.
However, the growth and fragmentation model remains the mono-disperse model \citep{stepinski_global_1997} in both cases. In particular, it cannot account for collisions between particules of very different sizes or masses. High-mass-ratio collisions have been found to produce stronger compression \citep{Tanaka_2023}. Taking them into account would help aggregates to reach their compacted state sooner. The mono-disperse model nonetheless results in global size distributions that are similar to those obtained with polydisperse models \citep{gonzalez_self-induced_2017,Vorobyov_2018}. The reader is referred to \citet{laibe_growth_2008,vericel_self-induced_2020,vericel_dust_2021,Michoulier_erosion} for additional discussions of this formalism.

\begin{table}
	\centering
	\caption{Summary of the classifications of cometary grains from the Rosetta and Stardust missions, adapted from \citet{guttler_synthesis_2019}.}
   \label{Tab:guttler_2019}
	\begin{tabular}{*{4}{c}}
		\hline\hline
		Group & $s$ & $\phi$ & Contribution\\
		\hline
        \shortstack{Porous\\ \, } & \shortstack{1 $\mu$m -- 1 m\\ \,} & \shortstack{$0.05$ to $0.9$\\ \,} & \shortstack{Dominates the\\ size distribution} \\
        Fractal& 1 $\mu$m -- 10 mm & $<0.05$ & Low fraction \\
        Solid & 0.1 $\mu$m -- 0.5 mm & $>0.9$ & Very common \\
		\hline
	\end{tabular}
\end{table}

\begin{figure*}
    \centering
    \includegraphics[width=\textwidth,clip,trim=.5cm 2.7cm .5cm 2cm]{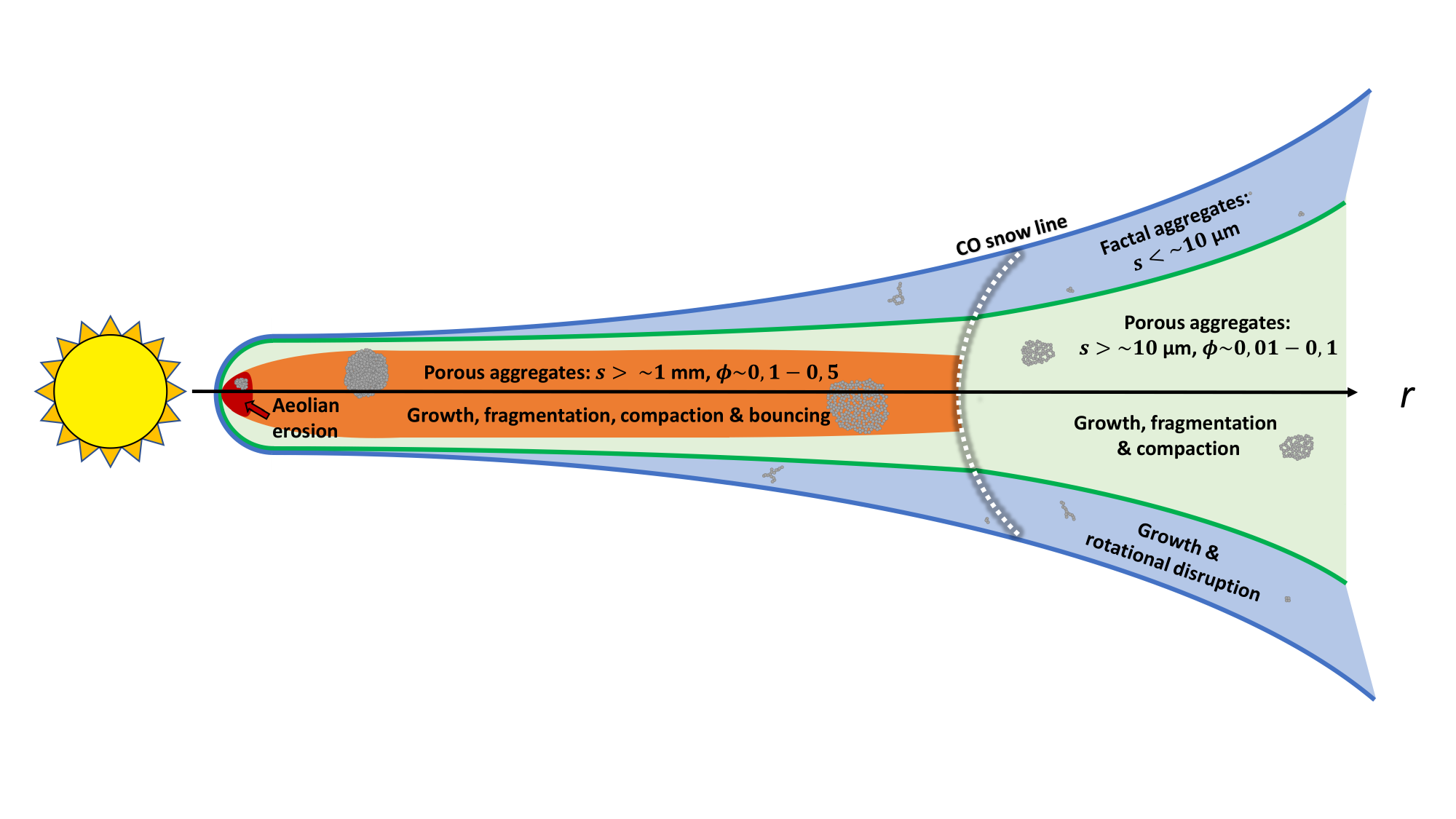}
    \caption{Diagram summarizing the different physical processes related to dust seen and studied in this paper.}
    \label{Fig:synthesis_draw}
\end{figure*}

The 3D simulations use the dust-as-mixture formalism in the terminal velocity approximation, which is no longer valid when dust decouples from the gas. Moreover, the dust-as-mixture (one fluid) simulations do not depict the formation of self-induced dust traps, whereas previous dust-as-particles simulations showed this mechanism \citep{Michoulier_SF2A_2022}. 
Future dust-as-particles simulations, albeit significantly more time-consuming, should verify the presence of these traps, as well as demonstrate whether dust can effectively decouple.

The fragmentation model is not entirely consistent. We set the fragmentation threshold to a 
predefined, fixed value. In practice, the fragmentation threshold should depend on the size and porosity 
of grains. \citet{garcia_evolution_2018} tested a model with a variable threshold with limited success. 
However, there is no existing model or study that establishes this relationship.
A consistent model is needed to better understand the fragmentation of aggregates and the its influence on the evolution of dust grains.

In a similar vein, the difficulty to obtain material properties for different compounds due to the complexity of experimental studies and the large parameter space they would need to cover makes assumptions unavoidable in order to develop a general model. Some of them may be lifted in the future as more data becomes available, for example measurements of the Young's modulus over a wide range of porosities for several species of astrophysical interest.

\section{Conclusions}
\label{Sc:Conclusions}

A key missing piece in our understanding of the puzzle of planet formation is how sub-micron-sized dust grains coagulate into planetesimals while avoiding the fragmentation and radial drift barriers.

These obstacles must be overcome for dust to survive long enough to form kilometre-sized rocky bodies.
While solutions have been proposed to explain planetesimal formation, mechanisms such as the streaming instability \citep{youdin_streaming_2005} involve specific conditions, such as the presence of large dust particles settled in the midplane. However, there is no clear answer on how to achieve these conditions. Growth up 
to a few microns is relatively straightforward, but reaching millimetre sized or larger grains, as observed, is difficult because of radial drift.

To address these issues, we explored the evolution of porous grains, including various mechanisms that destroy them. 
We developed a new physical model and code module to account for growth and porosity evolution, as well as bouncing and fragmentation with compaction, within the mono-disperse approximation.
We investigated the influence of porosity on dust evolution using the 1D code \textsc{Pamdeas} \citep{Michoulier_Gonzalez_Disruption} and the 3D SPH code \textsc{Phantom} \citep{price_phantom_2018}.
We performed simulations varying porosity, compaction, the constituent species of the grains, and the fragmentation threshold. Our conclusions are as follows:
\begin{enumerate}
\item We confirm that porosity enables the formation of millimetre to centimetre-sized grains \citep{garcia_evolution_2018}. The main limiting factor is the fragmentation threshold.
\item With growth and fragmentation alone, we find large values for the filling factor ($\phi \sim 10^{-4}$--$10^{-3}$) that are incompatible with observations, which show compact grains in the inner regions.
\item  When including compaction during fragmentation and bouncing, we formed grains of several hundred $\mu$m to a few mm, but this time with filling factors compatible with observations ($\phi \in [0.1;1]$).
\item 3D simulations with high turbulent viscosity showed signs of rotational disruption in the intermediate layers of the disc, when aggregates are settling in the outer regions, but we found the overall impact of rotational disruption on dust evolution to be small.
\end{enumerate}

Figure~\ref{Fig:synthesis_draw} summarises the key mechanisms involved in the evolution of grains.
Overall, we  found that considering porosity, compaction effects, and snow lines can reconcile observations with theoretical models of planet formation. Future work may involve developing a coherent model for the fragmentation threshold, further investigating the impact of snow lines, and using simulations to generate synthetic images to refine our understanding of observations.

\begin{acknowledgements}
We thank the anonymous referee for their report, which allowed us to improve the clarity of this paper.
The authors acknowledge funding from ANR (Agence Nationale de la Recherche) of France under contract number ANR-16-CE31-0013 
(Planet-Forming-Disks) and thank the LABEX Lyon Institute of Origins (ANR-10-LABX-0066) for its financial support within the Plan France 2030 of the French government operated by the ANR.
This research was partially supported by the Programme National de Physique Stellaire and the Programme National de Planétologie of CNRS (Centre National de la Recherche Scientifique)/INSU (Institut National des Sciences de l’Univers), France. We gratefully acknowledge support from the PSMN (Pôle Scientifique de Modélisation Numérique) of the ENS de Lyon for the computing resources. This project has received funding from the European Union's Horizon 2020 research and innovation programme under the Marie Sk\l{}odowska-Curie grant agreements No 210021 and No 823823 (DUSTBUSTERS). DJP is grateful for Australian Research Council Discovery Project funding via DP220103767 and DP240103290. We thank the Australian-French Association for Research and Innovation (AFRAN) for financial support.
Figures were made with the Python library \texttt{matplotlib} \citep{Hunter:2007}.
\end{acknowledgements}

%
%

\bibliographystyle{aa} 
\bibliography{porosity_evolution} 

\appendix

\section{Equations for porosity evolution during growth}
\label{App:Porosity_evol_growth}

\subsection{Hit \& Stick Regime}
\label{App:Porosity_evol_growth_h&s}

In the hit \& stick regime, the mass of the grain doubles during a collision, and a volume of void $V_\mathrm{void}$ is formed.
According to \citet{okuzumi_numerical_2009}, this captured volume $V_\mathrm{void}$ corresponds to $99\%$ of the initial volume 
$V_\mathrm{i}$, thus adding to the initial volumes of the two grains before the collision.
The filling factor $\phi\propto m/V$ is therefore multiplied by a factor of 2/2.99.
For an arbitrary number of collisions $n$, there is a geometric relationship between the initial filling factor 
$\phi_0$ and the final $\phi_\mathrm{f}$, given by $\phi_\mathrm{f} = (2/2.99)^n \phi_0$ \citep{garcia_evolution_2020}. Moreover, the mass of a grain after the same number 
of collisions has grown by $m_\mathrm{f} = 2^n m_0$, $m_0$ being the monomer mass. By isolating $n$ and combining the two expressions, we obtain the filling factor 
in the hit \& stick regime \citep[assuming $E_\mathrm{kin} \ll E_\mathrm{roll}$,][]{garcia_evolution_2020}
\begin{equation}\label{Eq:h&s}
    \phi_\mathrm{h\&s} = \left(\frac{m}{m_0}\right)^\psi,
\end{equation}
with $\psi=\ln\left({2/2.99}\right)/\ln{2}\simeq-0.58$, where for the monomer $\phi_0=1$.

\subsection{Collisional compression regime}
\label{App:Porosity_evol_growth_coll_comp}

As both the grain mass and the relative velocity $v_\mathrm{rel}$ increase, the kinetic energy increases.
Since $v_\mathrm{rel}$ depends on the Stokes number (Sect.~\ref{Ssc:Growth Model}), which in turn depends on the drag regime, there is an expression for each case \citep{garcia_evolution_2020}.
\citet{suyama_geometric_2012} provide the volume of a grain after collision $V_\mathrm{f}$ in terms of the volume of a 
grain before collision $V_\mathrm{i}$. This relation can be transformed to incorporate filling factors using the fact 
that $\phi_\mathrm{f} = m_\mathrm{f}/V_\mathrm{f}\rho_\mathrm{s} = 2 m_\mathrm{i}/V_\mathrm{f}\rho_\mathrm{s}$ \citep{garcia_evolution_2020}
\begin{equation}
    \phi_\mathrm{f} = \frac{2 m_\mathrm{i}}{\rho_\mathrm{s}} \left(\frac{(3/5)^5 \left(E_\mathrm{Kin} - 3b\,E_\mathrm{roll}\right)}{N_\mathrm{tot}^5 \, b \,E_\mathrm{roll} V_0^{10/3}} +\frac{V_\mathrm{i}^{-10/3}}{2^4}  \right)^{3/10}.
\end{equation}
Here, $N_\mathrm{tot}$ is the total number of monomers, which is $2 m_\mathrm{i}/m_0$, and $V_0$ is the volume of a monomer.
This model is recursive, meaning that the evolution of porosity is tracked after each collision. However, in the case 
of global simulations, the integration time step rarely aligns with the time interval between collisions $\tau_\mathrm{coll}$. 
Therefore, \citet{garcia_evolution_2020} transformed this discrete model into a continuous 
one to depend solely on the grain mass and quantities related to $r$, enabling its implementation in codes for global simulations.
To achieve this, they assumed the approximation $E_\mathrm{kin} \gg E_\mathrm{roll}$.

In the Epstein and Stokes regimes for $\mathrm{St}<1$, one obtains
\begin{align}
    \phi_\mathrm{Ep,St<1} &= \left(\frac{A}{2\left(2^{3/40}-1\right)}\right)^{3/8} \left(\frac{m}{m_0}\right)^{-1/8},\label{Eq:phi,EpSt<1}\\[10pt]
    \phi_\mathrm{St,St<1} &= \left(\frac{A c_\mathrm{g} a_0}{9 \nu_\mathrm{mol} \left(2^{1/5}-1\right) }\right)^{1/3}.
    \label{Eq:phi,StSt<1}
\end{align}
\begin{equation}
    \nu_\mathrm{mol}=\frac{5\sqrt{\pi}}{64}\frac{m_\mathrm{mol}c_\mathrm{g}}{\rho_\mathrm{g}\sigma_\mathrm{mol}},
    \label{Eq:nu_mol}
\end{equation}
is the gas molecular kinematic viscosity, where $m_\mathrm{mol}$ and $\sigma_\mathrm{mol}$ are the mass and molecular cross section of the H$_2$ molecule, and $A$ is a factor used to simplify the expressions and is given by
\begin{equation}
    A = \frac{243\sqrt{2}\pi}{15625}\frac{\mathrm{Ro} \,\alpha\, a_0^4 \,\rho_\mathrm{s}^2 c_\mathrm{g} \Omega_\mathrm{K}}{\rho \, b \, E_\mathrm{roll}},
\end{equation}
where $\rho=\rho_\mathrm{g}+\rho_\mathrm{d}$ is the total density.
It is observed that in the Stokes regime with $\mathrm{St} < 1$, the filling factor depends only on the local disc conditions.
For regimes with $\mathrm{St} > 1$, collisional compression is less efficient due to the decrease in $v_\mathrm{rel}$. 
By neglecting $E_\mathrm{kin}$, it can be derived from \citet{okuzumi_rapid_2012} and \citet{garcia_evolution_2020} that $\phi_\mathrm{f} \propto m_\mathrm{f}^{-1/5}$ for $\mathrm{St}>1$.
Therefore, the filling factors for $\mathrm{St}>1$ can be expressed as
\begin{align}
    \phi_\mathrm{Ep,St>1} &= \phi_\mathrm{Ep,St<1}(M_4) \left(\frac{m}{M_4}\right)^{-1/5},\\[10pt]
    \phi_\mathrm{St,St>1} &= \phi_\mathrm{St,St<1}(M_5) \left(\frac{m}{M_5}\right)^{-1/5},
\end{align}
where $M_4$ and $M_5$ are the transition masses between the Epstein and Stokes regimes for $\mathrm{St} < 1 $ and $\mathrm{St} > 1$.
The transition masses are given in Appendix~\ref{App:Transition_Masses}. The equations giving $\phi$ in the various regimes can also be rewritten as a function of the size $s$ instead of the mass $m$, they are listed in Appendix~\ref{App:phi_s}. However, the natural variable to describe dust evolution is $m$.

The filling factor resulting from the hit \& stick and collisional compression regimes is noted $\phi_\mathrm{coll}$.

\subsection{Static compression regime}
\label{App:Porosity_evol_growth_static_comp}

In addition to undergoing compression due to collisions, aggregates can also be statically compacted 
either by the surrounding gas or by their own gravity.
\citet{kataoka_static_comp_2013} and \citet{garcia_evolution_2020} provide the relationship between static 
compression pressure and the filling factor
\begin{equation}\label{Eq:Pcomp}
    \phi = \left(\frac{a_0^3 P_\mathrm{comp}}{E_\mathrm{roll}}\right)^{1/3},
\end{equation}
which is only valid for small filling factors $\phi<0.1$.
In the case of compression by gas, the pressure is given by:
\begin{equation}
    P_\mathrm{comp} = \frac{F_\mathrm{D}}{\pi s^2}=\frac{m \Delta v \Omega_\mathrm{K}}{\pi s^2 \mathrm{St}},
\end{equation}
which allows to obtain the expression for the filling factor $\phi_\mathrm{gas}$ 
\citep{garcia_evolution_2020}:
\begin{equation}
    \phi_\mathrm{gas}=\left(\frac{m_0 \,a_0}{\pi E_\mathrm{roll}}\frac{\Delta v \, \Omega_\mathrm{K}}{\mathrm{St}}\right)^{3/7}
    \left(\frac{m}{m_0}\right)^{1/7}.
\end{equation}
For self-gravity, the exerted pressure is given by
\begin{equation}
    P_\mathrm{comp} = \frac{\mathrm{G} m^2 }{\pi s^4},
\end{equation}
and the expression for the filling factor $\phi_\mathrm{grav}$ is \citep{garcia_evolution_2020}:
\begin{equation}
    \phi_\mathrm{grav}=\left(\frac{\mathrm{G} m_0^2}{\pi a_0 E_\mathrm{roll}}\right)^{3/5}
    \left(\frac{m}{m_0}\right)^{2/5},
\end{equation}
where G is the universal gravitational constant.
To obtain the final filling factor of a grain with mass $m$ at a distance $r$ from the star, we take the value of $\phi$ 
given by the phenomenon that compacts the grain the most, which is the largest value among static compression by gas, 
static compression by gravity, and compression by collision: $\phi_\mathrm{min}= \max(\phi_\mathrm{coll}$, $\phi_\mathrm{gas}$, $\phi_\mathrm{grav})$.
All these regimes are presented in Fig.~\ref{Fig:fig_croissance_pamdeas} in the ($s$,$\phi$) plane for different distances to the star.

\subsection{Growth from above the compression limit}
The minimum filling factor must be $\phi_\mathrm{min}$. However, there are cases where the evolution of a grain (e.g. via bouncing) can produce a filling factor larger than $\phi_\mathrm{min}$, placing it in the space above the curves showed in Fig.~\ref{Fig:fig_croissance_pamdeas}.
To take into account the increase of porosity of grains during growth above the limit, we can compute the filling factor as a function of the initial filling factor $\phi_\mathrm{i}$, and the final and initial masses $m_\mathrm{f}$ and $m_\mathrm{i}$ in the case where $v_\mathrm{rel}<v_\mathrm{frag}$.
\begin{align}
    \phi_\mathrm{grow} &= \phi_\mathrm{i} \left(\frac{m_\mathrm{f}}{m_\mathrm{i}}\right)^\psi\quad \text{if } E_\mathrm{kin} < 3 b E_\mathrm{roll},\\
    \phi_\mathrm{grow} &= \phi_\mathrm{i} \left(\frac{m_\mathrm{f}}{m_\mathrm{i}}\right)^{-1/5}\:\text{otherwise}.
\end{align}
The resulting filling factor is then the maximum of $\phi_\mathrm{grow}$ and $\phi_\mathrm{min}$.

\section{Transition masses}
\label{App:Transition_Masses}

\begin{figure}
    \centering
    \includegraphics[width=\columnwidth,clip]{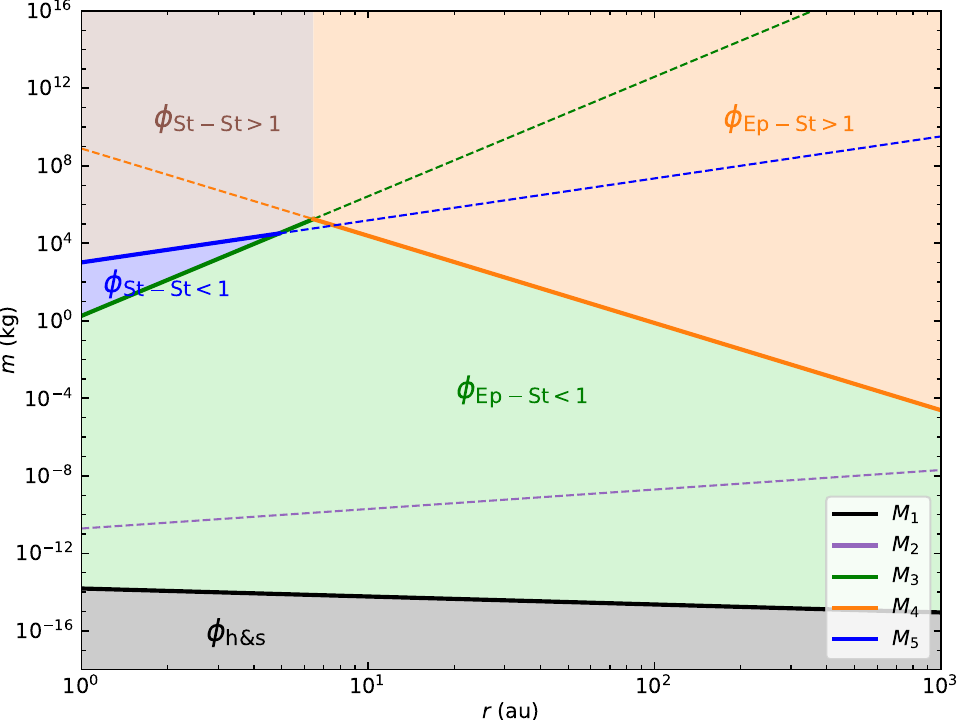}
    \caption{Transition masses for our standard disc model and for silicate grains.}
    \label{Fig:rMtr_STD_log}
\end{figure}

The transition masses are the masses at which the behavior of a grain changes. 
The different expressions of $\phi$ found for each regime in Sect.~\ref{Ssc:Porosity Model} are independent, the transition masses allow us to relate them.
For the transition masses $M_1$ and $M_2$ that connect the hit \& stick regime to the Epstein and Stokes regimes with $\mathrm{St} < 1$, we find
\begin{align}
    \frac{M_1}{m_0} &= \left(\frac{A}{2\left(2^{3/40}-1\right)}\right)^\frac{3/8}{\psi + 1/8},\\[10pt]
    \frac{M_2}{m_0} &= \left(\frac{A c_\mathrm{g} a_0}{9 \nu_\mathrm{mol} \left(2^{1/5}-1\right) }\right)^{1/3\psi}.
\end{align}
The transition mass $M_3$ corresponds to the transition between the Epstein and Stokes regimes
\begin{align}
    \frac{M_3}{m_0} &= \left(\frac{A}{2\left(2^{3/40}-1\right)}\right)^3 \left(\frac{A c_\mathrm{g} a_0}{9 \nu_\mathrm{mol} \left(2^{1/5}-1\right) }\right)^{-8/3}, \\
    &= \left(\frac{M_1}{m_0}\right)^{8\psi+1} \left(\frac{M_2}{m_0}\right)^{-8\psi}
\end{align}
Finally, the expressions for $M_4$ and $M_5$ are obtained by equating the expression of $\phi$ for the Epstein or Stokes 
regimes with $\mathrm{St} < 1$ to the expression of $\phi$ obtained by setting $\mathrm{St}=1$.
\begin{align}
    \frac{M_4}{m_0} &= \left(\frac{\rho c_\mathrm{g}}{\rho_\mathrm{s} a_0 \Omega_\mathrm{K}}\right)^4 \left(\frac{A}{2\left(2^{3/40}-1\right)}\right)^{-1},\\[10pt]
    \frac{M_5}{m_0} &= \left(\frac{9 \nu_\mathrm{mol} \rho}{2\rho_\mathrm{s} a_0^2 \Omega_\mathrm{K}}\right)^{3/2} \left(\frac{A c_\mathrm{g} a_0}{9 \nu_\mathrm{mol} \left(2^{1/5}-1\right) }\right)^{-1/6}.
\end{align}

Here, we have not considered the other Stokes regimes that are reached for even larger sizes than the linear regime as static grain compression is always dominant compared to these regimes.
The various transition masses as well as the different expansion and compression regimes are shown in Fig.~\ref{Fig:rMtr_STD_log}, inspired by Fig.~A1 of \citet{garcia_evolution_2020}.
Note that the transition mass $M_2$ is always larger than mass $M_1$. The Stokes regimes are reached in the inner regions of a disc, at a distance less than 10 au.

\section{Filling factor as a function of size}
\label{App:phi_s}

The equations giving the filling factor $\phi$ as a function of mass $m$ in Appendix~\ref{App:Porosity_evol_growth} can be rewritten as a function of the size $s$ as

\begin{equation}\label{Eq:phi_hs_size}
     \phi_{\mathrm{h\&s}}=\left(\frac{s}{a_0}\right)^{\frac{3\,\psi}{1-\psi}},
\end{equation}

\begin{align}
\phi_{\mathrm{Ep-St}<1}&=\left(\frac{A}{2(2^{3/40}-1)}\right)^{1/3}  \left(\frac{s}{a_0}\right)^{-1/3}\label{eq:ep,st<1(s)},\\[10pt]
\phi_{\mathrm{St-St}<1}&=\left(\frac{A\,c_{\mathrm{g}}\, a_0}{9\nu(2^{1/5}-1)}\right)^{1/3},\label{st:st,st<1(s)}
\end{align}

\begin{align}
    \phi_{\mathrm{Ep-St}>1}&=\left(\frac{\rho\,c_{\mathrm{g}}}{\rho_{\mathrm{s}}\, \Omega_{\mathrm{K}}\,a_0}\frac{A}{2^{3/40}-1}\right)^{1/4} \left(\frac{s}{a_0}\right)^{-1/2},\label{eq:ep,st>1(s)_2}\\[10pt]
    \phi_{\mathrm{St-St}>1}&=\left(\frac{\rho\,c_{\mathrm{g}}\, a_0}{2\rho_{\mathrm{s}} \,\Omega_{\mathrm{K}} \,a_0^2}\,\,\frac{A\,\, }{2^{1/5}-1}\right)^{1/4} \left(\frac{s}{a_0}\right)^{-1/2}.\label{eq:st,st>1(s)_2}
\end{align}

\section{Co-evolution of grain size and filling factor in \textsc{Phantom} simulations with and without compaction}
\label{App:Phantom-Compaction-effect-s-phi-r}

\begin{figure*}
  \centering
  \includegraphics[width=\textwidth,clip]{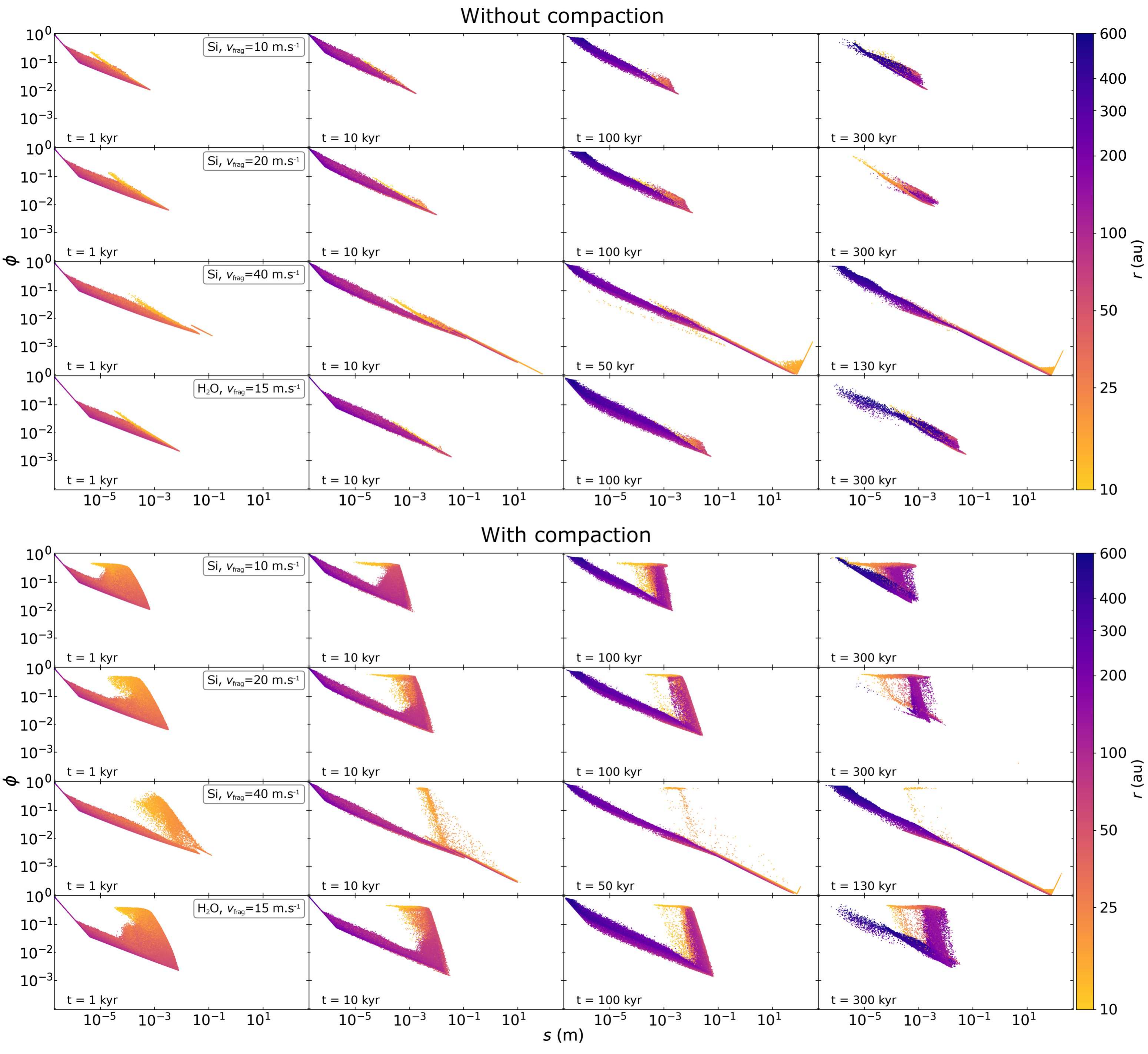}
  \caption{Same as Fig.~\ref{Fig:Phantom-Compaction-effect-r-s-phi} but showing the filling factor $\phi$ plotted against grain size, with colour indicating the distance $r$.}
  \label{Fig:Phantom-Compaction-effect-s-phi-r}
\end{figure*}

Fig.~\ref{Fig:Phantom-Compaction-effect-s-phi-r} shows the joint evolution of grain size and filling factor in \textsc{Phantom} simulations without (top) and with (bottom) compaction during fragmentation and bouncing, with the same order of simulations as in Fig.~\ref{Fig:Phantom-Compaction-effect-r-s-phi}. Without compaction (top panel), the trajectories for all simulations are very similar to the 1D case (Fig.~\ref{Fig:Pamdeas-Compaction-effect-r-s-phi}), with a spread corresponding to the disc's vertical extent. When compaction is considered (bottom panels), the compaction trajectory is more vertical compared to that obtained with \textsc{Pamdeas}, and the bouncing plateau at $\phi \sim 0.4$, primarily populated with grains very close to the disc inner edge, is more clearly visible. The overall evolution appears more complex and the $s-\phi$ plane is more densely populated. A large spread of filling factors can be reached for a given grain size and at a given distance, grains of various $s$ and $\phi$ can be found, indicating the coexistence of porous and compacted dust.

Simulations with silicate grains and $v_\mathrm{frag,\:Si} = 10$ and 20~$\msec$ (first two rows) are again quite similar, with grains reaching larger sizes and porosities before being fragmented and compacted for the latter value. The main difference is that at $t=300$~kyr almost all grains are compacted or undergoing compaction for $v_\mathrm{frag,\:Si} = 20$~$\msec$, with very few left on the growth branch, while for 10~$\msec$ a large fraction of grains are still growing, mainly in the outer disc. Despite the fact that simulations GF-Si-a02-Vf40 and GBFc-Si-a02-Vf40 (third rows) are not valid, it is still interesting to note that, due to larger sizes being reached, the gas and self-gravity compaction regimes are visible for solids larger than a few m (see Fig.~\ref{Fig:fig_croissance_pamdeas}).
Again, the water ice simulations (bottom rows) are similar to that with silicate grains and $v_\mathrm{frag,\:Si} = 20$~$\msec$, with larger $s$ and smaller $\phi$, except for a fraction of still growing grains at large $r$.

\section{Influence of rotational disruption}
\label{App:influence_of_rotational_disruption}

\begin{figure*}
    \centering
    \includegraphics[width=\textwidth,clip]{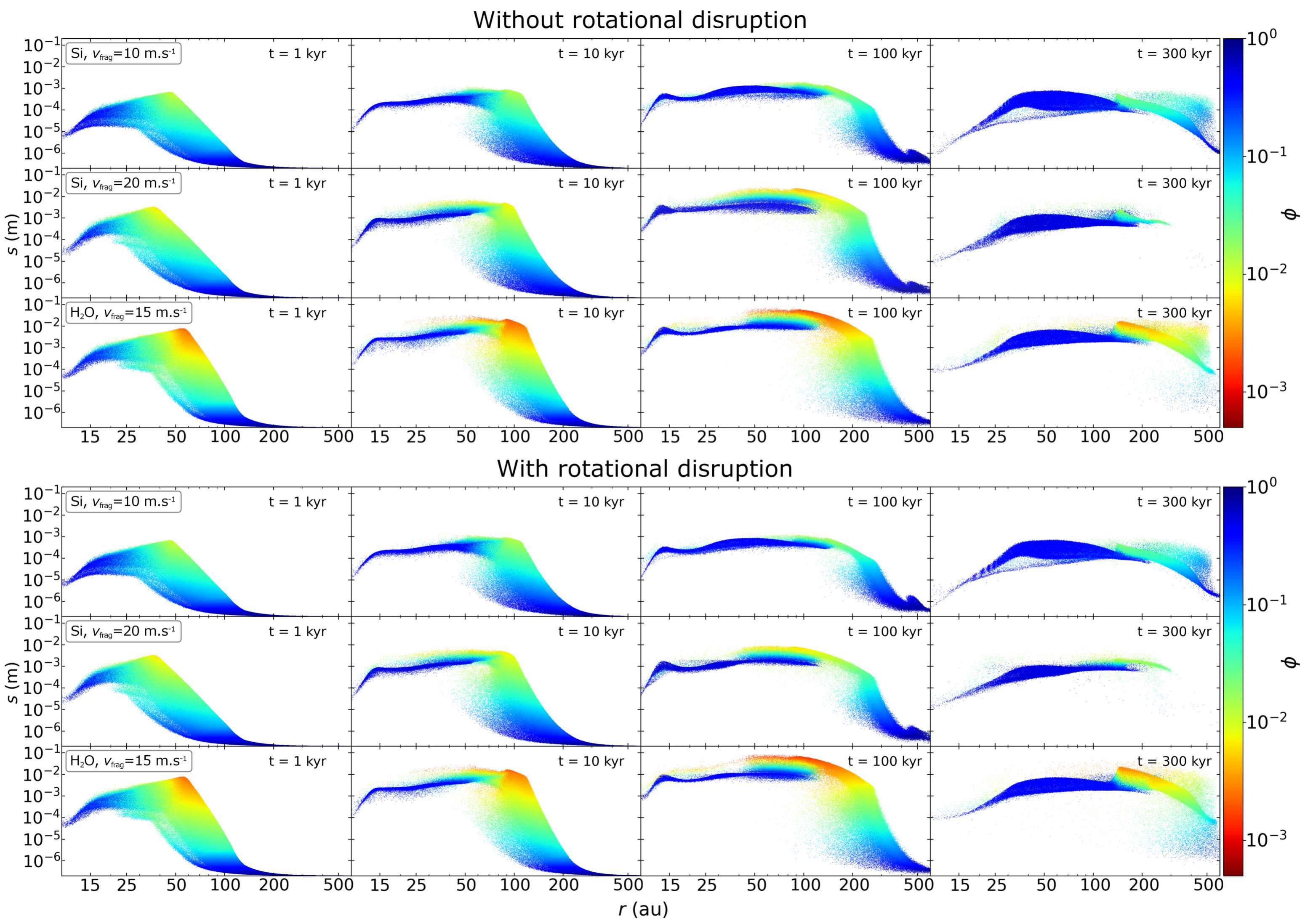}
    \caption{Comparison between \textsc{Phantom} simulations with growth, bouncing, fragmentation with compaction, and without (GBFc-*-a02-*, top panel) and with (GBFcD-*-a02-*, bottom panel) rotational disruption showing the radial grain distribution, with colour representing the dust filling factor. Simulation parameters are indicated in each row.}
    \label{Fig:Phantom-disruption-effect-r-s-dtgr}
\end{figure*}

\citet{Michoulier_Gonzalez_Disruption} studied the effect of rotational disruption on the evolution of porous grains using 1D simulations with \textsc{Pamdeas}. 3D disc simulations taking into account the evolution of gas and dust are necessary to understand in more detail when and where rotational disruption influences dust evolution. To that effect, we performed simulations modelling this phenomenon for silicates with $v_\mathrm{frag,\:Si} = 10$ and 20~$\msec$, and for water ice with $v_\mathrm{frag,\:Si} = 15$~$\msec$, listed in the last three rows of Table~\ref{tab:simulations_names}. The grain properties are somewhat different compared to \citet{Michoulier_Gonzalez_Disruption}: the monomer size is set to 0.2 instead of 0.1~$\mu$m, and $\gamma_\mathrm{s} = 0.2\ \jmsqare$ for silicates. In fact, with these new parameters, the tensile strength $S_\mathrm{max}$ should be larger by a factor of two.

Simulations with \textsc{Pamdeas} showed signs of rotational disruption only for very small values of the viscosity parameter $\alpha < 5\times10^{-4}$, lower than that used with \textsc{Phantom}. The conditions necessary for its appearance in the midplane (which \textsc{Pamdeas} simulates) are therefore challenging to achieve.

Figure~\ref{Fig:Phantom-disruption-effect-r-s-dtgr} compares 3D simulations including growth, bouncing and fragmentation with compaction, without (top) and with (bottom) rotational disruption. It shows the radial grain distribution, with colour representing the dust filling factor $\phi$. For simulations GBFc-Si-a02-Vf10 and GBFcD-Si-a02-Vf10 (top rows in both panels), very few differences are observed, which is expected since the fragmentation threshold is low enough for grains to fragment before they can be rotationally disrupted (which only happens at higher relative velocities and, therefore, larger sizes).
For simulations GBFc-Si-a02-Vf20 and GBFcD-Si-a02-Vf20 (middle rows), rotational disruption is able to destroy grains during their fall towards the midplane, which slows down their settling and limits dust 
enrichment in the midplane. Since the dust density is slightly lower, grains do not grow as much as in the case without disruption. This is mostly visible for grains outside of 50~au at $t=100\ \mathrm{kyr}$. After 300~kyr, the differences have been largely erased by the subsequent dust evolution.
Finally, for simulations GBFc-H2O-a02-Vf15 and GBFcD-H2O-a02-Vf15 (bottom rows), disruption has a minimal effect on the size and porosity of dust grains, only visible outside of 150~au. Indeed, the growth of water ice grains is very rapid and largely counteracts rotational disruption. It is at $t=300\ \mathrm{kyr}$ that the effect becomes more visible since, unlike the simulation without disruption, many small grains are observed between 200 and 500 au.

In general, disruption has little effect on the size and porosity of grains and their evolution in the disc, with fragmentation and compaction largely dominating grain growth and destruction.

\end{document}